\documentstyle[11pt,psfig,rotating,axodraw]{article}

\def\NIM{{\em Nucl. Instrum. Methods}}

\def\NP{{\em Nucl. Phys.}}

\def\PL{{\em Phys. Lett.}  }
\def\PRL{\em Phys. Rev. Lett.}
\def\PR{{\em Phys. Rev.} }  
\def\ZP{{\em Z. Phys.} }

\def\APP{{\em Acta Physica Polonica} }

\def\PRL{{\em Phys. Rev. Lett.}}

\def\PRep{{\em Phys. Rep.}}

\def\EPJ{{\em Eur. Phys. Journal} }
\def\ARNS{{\em Annu. Rev. Nucl. Sci.}}

\def\kbn{Polish Committee for Scientific Research 2P03B01414.}

\def\ra{\rightarrow}

\def\tb{\tan \beta}

\def\be{\begin{equation}}
\def\ee{\end{equation}}
\def\bea{\begin{eqnarray}}
\def\eea{\end{eqnarray}}
\def\ie.{{$ i.e.$}}
\def\eg.{{$ e.g.$}}

\begin{document}
\title{
\begin{flushright}
IFT 98/17\\[1.5ex]
{\large \bf hep-ph/9812493} \\
\end{flushright}
~~~\\
~~~\\
{\Large{\bf {WHERE IS THE HIGGS BOSON?}}}
{\thanks{Presented at the XXXVIII Cracow School of
 Theoretical Physics, Zakopane, Poland, June 1-10, 1998.}\ }
{\thanks{Supported in part by \kbn}}}
 
\author{{\large {Maria Krawczyk}} \\
Institute of Theoretical Physics, 
Warsaw University,\\
 ul. Ho\.za 69,
00-681 Warsaw, Poland}

\maketitle
\begin{abstract}
The status of the Higgs boson search in the Standard Model and beyond it
is presented.
\end{abstract}

\section{Introduction}

There are numerous 
"elementary 
particles" in nature, Review of Particle Properties AD 1998
 lists about 1000  such 
particles  \cite{pdp98}.
The Standard Model describing these various particles together with 
 their interactions 
introduces far less numerous fundamental objects:\\
$$
\begin{array}{|c|c|c|c|c|}
\hline
spin ~{{1}\over{2}}\hbar& leptons:&\nu_{e}&\nu_{\mu}&\nu_{\tau}\\
&~~&e&\mu&\tau\\
&quarks:&u&c&t\\
&&d&s&b\\
\hline
spin ~1 \hbar&gauge ~~bosons:&\multicolumn{3}{|c|}{\gamma,~W,~Z,~g}\\
\hline
\end{array}
$$
where 
 quarks and gluons appear in, respectively,  three and eight states with
 different  colors, and  in addition the
 corresponding antiparticles should be included.

It is  difficult 
to imagine that also in future  all these particles
 remain   fundamental building blocks of matter. 
The appearance of fermions in the
three generations has no explanation so far, 
masses of fundamental particles 
 span a vast range  from zero for  the  photon and the gluon to 
80/90 GeV for  other two gauge bosons $W/Z$, and  
from a few MeV to  up 175 GeV for quarks ($u$, $d$ and $t$), 
finally for leptons from 0.5 MeV ($e$)
to 1.8 GeV (for $\tau$).  Do we understand this picture?

Today the most promissing  attempt
to establish the origin of mass both for
gauge bosons and fermions, and to explain  the  mass pattern, 
is based on
 {\sl the idea that all particles and 
forces started out as massless quanta but somehow acquired their various
masses because of their mutual interactions \cite{nambu}. }
It looks like a paradox that the key
to the  mass, being of a dynamical nature according to this conjecture,
 lies in the symmetry. 

Two kinds of symmetries are studied in particle physics.
From early days of particle physics  a symmetry considerations
were successfully applied
to describe  static properties of particles. To this purpose
the {\underline {global}} symmetries were involved. 
{\sl A {{global}} symmetry ... merely happens to exist, 
there seems to be no 
compelling  reason to think that things could not be other than they are.
A {\underline {local}} symmetry has a much more exalted status,
 because it is intimately 
connected with the basic forces of nature \cite{nambu}.}
Indeed
the application  of the  local invariance or symmetry
(gauge) principles
to describe the interactions   turned out to be  
one of the main achievement in the theory of elementary particles
in the  last 30 years.

The Standard Model, the theory  which describes electromagnetic,
 weak and strong 
interactions is based on the local symmetry  of the following form:
$$SU(2)_{I_W}\times U(1)_{Y_W} \times SU(3)_c,$$
where the local (gauge) groups   act
  on the      weak isospin ${I_W}$, the weak hypercharge  ${Y_W}$
and  the  color $c$ relevant for the strong interactions. 
The electric charge $Q$ is related to the above ``weak charges''
as follows:
$$
Q=I_W+{{Y_W}\over{2}}.
$$

Imposing the gauge symmetry  we agree (temporarily!) to  have 
$W/Z$ gauge bosons mediating the electro-weak interactions  
to be  massless and the fermions' masses
 will  be zero as well (the corresponding mass terms 
in the Lagrangian density would violate 
the assumed  symmetry).

The way to obtain a mass of gauge bosons $W/Z$,
 while preserving a local gauge symmetry, leads  to the concept of the 
spontaneous symmetry breaking (or {\sl the hidden symmetry}),
 which does not rely on the 
additional mass terms in the Lagrangian, but rather  
on the assumption that there exists a (scalar) field 
with a specific  form of interaction,  responsible for the mass
of  all the particles.
{\sl Mass then appears 
not as a result of emission and absorption of quanta of the scalar
field, but as a result of the interaction with the classical part
of the scalar field, which extends over all space \cite{vanstein}}.


The spontaneous symmetry breaking,  called  the  Higgs mechanism if applied  
to the local symmetry,
is considered as an 'origin' of the mass of fermions and gauge bosons 
in the Standard Model. The existence of the Higgs scalar 
is expected to be the direct physical manifestation of this mechanism.
Looking for a Higgs boson is therefore a challenge for 
the particle physics.
Of course it is not obvious whether the source of mass 
lies  within
the Standard Model. Much more probably it  
requires a wider scenario   beyond SM, presumably related to the 
unification of all fundamental forces (together with the gravity?).
\section{Spontaneous symmetry breaking. Higgs mechanism}

Let us consider  a simple example showing  how the  Higgs  mechanism works for 
a  system containing a gauge boson $A^{\mu}$.
 Here one introduces 
one  complex  scalar boson field  $\Phi$.  
  The interaction with the gauge boson is described by 
the   Lagrangian density with a local gauge group U(1) in the  following form: 
\begin{eqnarray}
{\cal L} =(D_{\mu}\Phi)(D^{\mu}\Phi)^*
+{{\mu^2}}\Phi^*\Phi
-{{\lambda}}(\Phi^*\Phi)^2-{{1\over4}}F^{\mu\nu}F_{\mu\nu},
\end{eqnarray}
where
\be
F^{ \mu \nu}=\partial^{\mu}A^{\nu}-\partial^{\nu}A^{\mu}.
\ee
The covariant derivative
\be
D^{\mu}=\partial^{\mu}+igA^{\mu}
\ee
  contains the term related to the
interaction  between the scalar and the gauge field with a coupling $g$. 
The considered Lagrangian density  is manifestly
symmetric 
under the local U(1) symmetry transformation. 

The  parameters in the potential part: 
\be
V=  -{{\mu^2}}\Phi^*\Phi
+{{\lambda}}(\Phi^*\Phi)^2,
\ee
are both positive,
\begin{equation}
\mu^2>0 \hspace{0.5cm}{\rm and~} ~\lambda>0,
\end{equation}
leading to the  potential bounded from below. (Note that negative $\mu^2$
 would  correspond to the conventional mass term for $\Phi$.)
The potential (4) has  minima for 
\begin{figure}
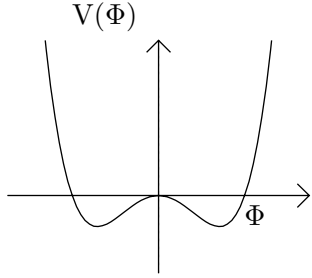

\hskip 1.0in
\input pot2.tex
\caption{The Higgs potential.}
\label{fig:Potential}
\end{figure}
 
\begin{equation}
|<\Phi>|={{1}\over{\sqrt{2}}}v=\sqrt{{\mu^2}\over{2 \lambda}}, 
\end{equation}
where $v$ is called a vacuum expectation value.
The example is shown in Fig.~\ref{fig:Potential}
  for the one real field $\phi$,
where two minima appear.  
By choosing one of these  minima as a a true minimum of the energy,
  the symmetry of the 
physical system  (the lowest (the vacuum) and higher excited states) 
is {\sl spontaneously broken}.

To make   the physical content of the theory 
more transparent the original field can be expressed by  new real fields,
 $\xi$ and $h$, with a zero
vacuum expectation values, as in the standard quantum field approach:
\begin{equation}
\Phi(x)={{e^{i\xi/v}}\over{\sqrt{2}}}(v+h(x)),
\end{equation}
and further by choosing a particular gauge with $\xi$=0 we get 
\begin{eqnarray}
{\cal L} ={{1}\over{2}}(\partial_{\mu}-igA_{\mu})
({{v}}+h)(\partial^{\mu}+igA^{\mu})
({{v}}+h)\\ \nonumber
+{{\mu^2}\over{2}}({{v}}+h)^2
-{{\lambda}\over{4}}({{v}}+h)^4-{{1}\over{4}}
F^{\mu\nu}F_{\mu\nu}=\\ \nonumber
{{1}\over{2}}(\partial_{\mu}h)(\partial^{\mu} h)-
\mu^2~h h+{{(gv)^{2}}\over{2}}A^{\mu}A_{\mu}+g^2 v  ~h A_{\mu}A^{\mu} +..
\end{eqnarray}

Interpreting the individual terms in the  Lagrangian density ${\cal L}$
one can find that the considered   theory contains: \\

\noindent
$\bullet$ a mass term for the gauge boson $M={{gv}}$,\\
$\bullet$ a neutral scalar boson $h$ (a real field) with a mass 
$\sqrt{2}\mu$,\\
$\bullet$ the interaction terms   $g M ~hA^{\mu}A_{\mu}$
 with the  coupling
proportional to  the mass of the gauge boson,\\ 
$\bullet$ the self interaction terms
$hhh$, $hhhh$ etc.\\

By  measuring the gauge boson mass one can determine the parameter 
$v$, provided there is independent constraint on the coupling $g$:
\be
M={{gv}}. 
\ee
However to obtain the mass of the Higgs boson we should know 
in the addition the self interaction, \ie.  parameter $\lambda$, since
\be
M_{h}=\sqrt{2 \lambda} v.                       
\ee
Note that massless gauge boson has only two polarization states, so by 
adding one complex or equivalently 
two real fields  we obtain enough 
independent components to describe one massive gauge boson with 
3 polarization
states and one neutral  scalar particle  $h$.
\section {Standard Model}
We  now summarize the situation of the Standard Model,
actually of its  electroweak sector (EW), where left-handed states of fermions 
constitute  the $SU(2)_{I_W}$ doublets, \eg. 
$
\left( \begin{array}{c}
\nu_{e} \\ e
\end{array} \right)_L
$, 
while the right-handed states are SU(2)$_{I_W}$ singlets. 
The neutrinos are massless in the standard version of the
Standard Model, and the corresponding right-handed states are missing.
 
To generate masses of weak bosons,
 one introduces one  complex scalar $SU(2)_{I_W}$ doublet (with $Y_W=1$)
 \cite{vanstein,hunter}: 
$$
 \Phi= \left( \begin{array}{c}
\phi^+ \\ \phi^0
\end{array} \right),
$$
which if a breakdown   in the form 
 SU(2)$\times$U(1) $\ra$ U$_{em}$ is assumed, 
the vacuum carries the quantum numbers of the neutral component,
 $<\phi^0>={{v}\over{\sqrt 2}}$.
The electroweak sector in SM  is described by a covariant derivative:
\begin{equation}
D^{ \mu}=\partial^{\mu}+ig{{\vec  { \tau}} \over{2}}{ \vec {W^{\mu}}} 
+ig'{{Y_W}\over{2}}B^{\mu},
\end{equation}
where the ratio of couplings $g$ and $g'$ is described by  
 the Weinberg angle $\theta_W$,
$\tan \theta_W=g'/g$.
The original vector gauge fields: 
\begin{equation}
W^{\mu}_1,W^{\mu}_2 {\rm {~and}} ~W^{\mu}_3,B^{\mu}  
\end{equation}
after mixing between the neutral fields, 
lead to the following physical  charged and neutral fields 
\begin{equation}
W_{\mu}^+,W_{\mu}^- {\rm {~and}} ~Z_{\mu}, A_{\mu},  
\end{equation}
with  the  corresponding  particles 
known as  $W^{\pm}$, $Z$ bosons and the photon, $\gamma$.
They mediate the so called charged (CC), neutral current (NC)
processes  and 
electromagnetic processes, respectively.

The mass formula for the $W$ boson is
as follows
\be
 M_W={{gv}\over {2}},
\ee 
 and therefore the Fermi constant $G_F$,
 measuring at low energy the strength of the CC weak interaction,
  can be directly 
related to the vacuum expectation value, giving
\begin{equation} 
G_F={{g^2}\over{4\sqrt2 M_W^2}} 
\ra v=(\sqrt2 G_F)^{-{1\over2}}=246 {\rm ~{GeV}}.
\end{equation}

Note that  generation of the masses for gauge bosons through 
 the Higgs mechanism
is such, that a simple relation between the gauge boson masses holds:
\begin{equation}
\rho={{M_W^2}\over{M_Z^2\cos^2\theta_W}}=1.
\end{equation}

In addition to  giving masses to the gauge bosons,
the interaction of scalar field 
leads to masses of  fermions. Namely for $f_L$ being a SU(2) doublet
and $f_R$ -- a SU(2) singlet  we get a mass term for the fermion $f$
\begin{equation}
g_f[(\bar f_{L}\Phi)f_{R}+hc]\ra 
{{g_f v}\over{\sqrt 2}}(\bar f f).
\end{equation}
Here $g_f$ is the so called  
Yukawa coupling for fermion $f$.                   
The parameter in front of the bracket in (20) for obvious reason should 
be interpreted as 
a mass of a fermion $f$, and therefore
\be
m_f={{{g_f}{v}}\over{\sqrt{2}}}.
\ee
 We conclude that
 the scalar field generates  mass terms 
for fermions. However we note that the fermions' 
masses are not fixed by the parameters of the Higgs potential,
nor the fermion mass pattern can be driven from the assumed mechanism.

The above expression  shows also the way  the scalar field
couples to fermions, with 
\be 
g_f={{\sqrt 2 m_f}\over {v}},
\ee
i.e. with a coupling proportional to the fermion mass, similarly 
as for the  Higgs boson coupling to 
the gauge bosons.

\section{A need for a Higgs boson in SM: a high energy limit}
Let us show now a different  argument for introducing a new particle
(a scalar) into the SM spectrum of fundamental particles, with couplings
 to gauge bosons and fermions precisely as discussed above for 
the Higgs boson.
The argument is based on a requirement that 
amplitudes for processes calculated in perturbation theory
should not   grow   too fast at high energy.
Following authors of \cite{vanstein} let us consider the  process
\begin{equation}
e^+e^-\ra \gamma\gamma
\end{equation}
 at very high energy.
The differential cross section describes this process 
as a function of the CM energy
$\sqrt s$ and the squared momentum transfer $t$, and
\begin{equation}
{{d\sigma}\over{dt}}\sim{{\alpha^2}\over{s^2}} ~~~{{\rm for}}~~s\sim|t|,
\end{equation}
which holds for the $s\ra \infty$.  Here $\alpha$ is the 
fine structure constant.

Let us now consider the corresponding process involving $Z$ bosons:
\begin{equation}
e^+e^-\ra ZZ.
\end{equation}
Should one expect for very large energy 
 $\sqrt s\gg M_Z$  
the same kind of behaviour as for photons (24)?

First we  observe that for the massive boson we have to
 include not only transverse but also
longitudinal polarization. Therefore  when calculating 
a cross section
the following sum over polarization states for $Z$ appears: 
\be
\sum{ {\epsilon}}_{\mu}{\epsilon^*_{\nu}}=-(g_{\mu\nu}-
{{k_{\mu}k_{\nu}}\over{M_Z^2}}),
\ee
with the high energy limit for the longitudinal polarization term  
\be
{{k_{\mu}}\over{M_Z}}\sim{{k_{0}}\over{M_Z}}
\ra \infty { \rm ~~~{for}} ~~~k_0\ra 
\infty ,
\ee
where $k_0$ is the $Z$ energy.
\begin{figure}[ht]
\center
\begin{tabular}{lcr}
\begin{picture}(105,45)(0,0)
\Line(40,20)(40,50)
\Text(0,15)[]{\mbox{$e^-$}}
\Text(0,50)[]{\mbox{$e^+$}}
\Text(85,10)[]{\mbox{$Z(\gamma)$}}
\Text(85,50)[]{\mbox{$Z(\gamma)$}}
\Vertex(40,20){2}
\Line(10,20)(40,20)
\Line(10,50)(40,50)
\Vertex(40,50){2}
\ZigZag(40,20)(70,20){3}{3}
\ZigZag(40,50)(70,50){3}{3}
\end{picture}
&
\begin{picture}(105,45)(0,0)
\Line(40,20)(40,50)
\Text(0,15)[]{\mbox{$e^-$}}
\Text(0,50)[]{\mbox{$e^+$}}
\Text(85,10)[]{\mbox{$Z(\gamma)$}}
\Text(85,50)[]{\mbox{$Z(\gamma)$}}
\Vertex(40,20){2}
\Line(10,20)(40,20)
\Line(10,50)(40,50)
\Vertex(40,50){2}
\ZigZag(40,20)(70,50){3}{3}
\ZigZag(40,50)(70,20){3}{3}
\end{picture}
&
\begin{picture}(105,45)(0,0)
\Line(10,20)(40,35)
\Line(10,50)(40,35)
\Text(0,15)[]{\mbox{$e^-$}}
\Text(0,50)[]{\mbox{$e^+$}}
\Text(95,10)[]{\mbox{$Z$}}
\Text(95,50)[]{\mbox{$Z$}}
\Vertex(40,35){2}
\DashLine(40,35)(65,35){4}
\Text(50,20)[]{\mbox{$Higgs$}}
\Vertex(65,35){2}
\ZigZag(65,35)(90,50){3}{3}
\ZigZag(65,35)(90,20){3}{3}
\end{picture}
\end{tabular}
\caption{The process $e^+e^-\ra \gamma \gamma$,
$e^+e^-\ra Z Z$ and the Higgs boson contribution to 
$e^+e^-\ra Z Z$.}
\label{fig:highen}
\end{figure}
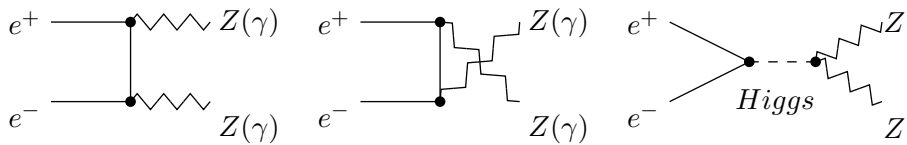
This behaviour 
leads to  the following high energy behaviour
for the cross section 
\begin{equation}
{{d\sigma}\over{dt}}\sim{{(g^2+g'^2)^2}\over{s^2}} 
{{m_e^2s}\over{M_Z^4}},
\end{equation}
with  the relevant  couplings $g$ and $g'$.
This growth with  energy is not accepted from point of view of 
unitarity.
 One can regularize this behaviour 
by adding to the Lagrangian density the contribution due to
the scalar particle $H$ interaction  in the form
\begin{equation}
{\cal L}_H=c_eH\bar e e +c_Z H Z^{\mu}Z_{\mu}.
\end{equation}
This leads to the extra contribution to the $ZZ$ production, 
see Fig.~\ref{fig:highen},
with couplings $c_e, ~c_Z$  adjusted to cancel 
the term  (25), namely:
\bea
c_ec_Z={1\over2}(g^2+g'^2)m_e,\\
c_e={{m_e}\over{v}},\\
c_Z={{g M_Z}\over{\cos\theta_W}}={{{{(g^2+g'^2)}}v}\over{2}}.
\eea

Considering 
 other  processes of these types  one  can obtain  couplings of scalar 
(Higgs) boson to other particles, 
all being proportional to the mass of the respective particle.

\section{Search for the Higgs particle in SM}
The Higgs particle is  the  only missing particle 
of  the Standard Model. All properties of this particle but its mass
are fixed by the Higgs potential. 
Theoretical constraints 
 from vacuum stability  place lower limit for the Higgs mass 
 $\sim$ 130 GeV while
the requirement of applicability of the perturbation theory
up to the unification scale, approximately $ 10^{16}$ GeV,
  gives upper  limit $\sim$ 180 GeV 
 \cite{hollik,vac}.

But  experiment remains the  decisive source of information on the 
spontaneous symmetry breaking.  
Direct searches, where one looks for
a possible production of the Higgs particle, 
 are being performed in $e^+e^-$
colliders : LEP and SLC at the $Z$ peak and above, 
in $p \bar p$ collider TEVATRON with a
 CM energy 1.8 TeV.
In these experiments  
 the Higgs boson with mass up to 107 GeV \cite{desch} and 120 GeV 
\cite{carzer,futtev},   respectively, can be found.
In the near future  the $pp$ collider LHC
will operate at 14 TeV with the mass coverage up to $\sim$ 1 TeV \cite{lhc}.
 There are plans to build $e^+e^-$, $e\gamma$ and
 $\gamma \gamma$  Linear Colliders (LC) \cite{lc}, 
 and maybe also Muon Colliders \cite{mu},
 where precise measurements of the properties of the SM Higgs boson
 can be performed.

\begin{figure}[hbtp]
\begin{center}
\hspace*{-6.5cm}
\begin{turn}{270}%
\epsfxsize=5.5cm \epsfbox{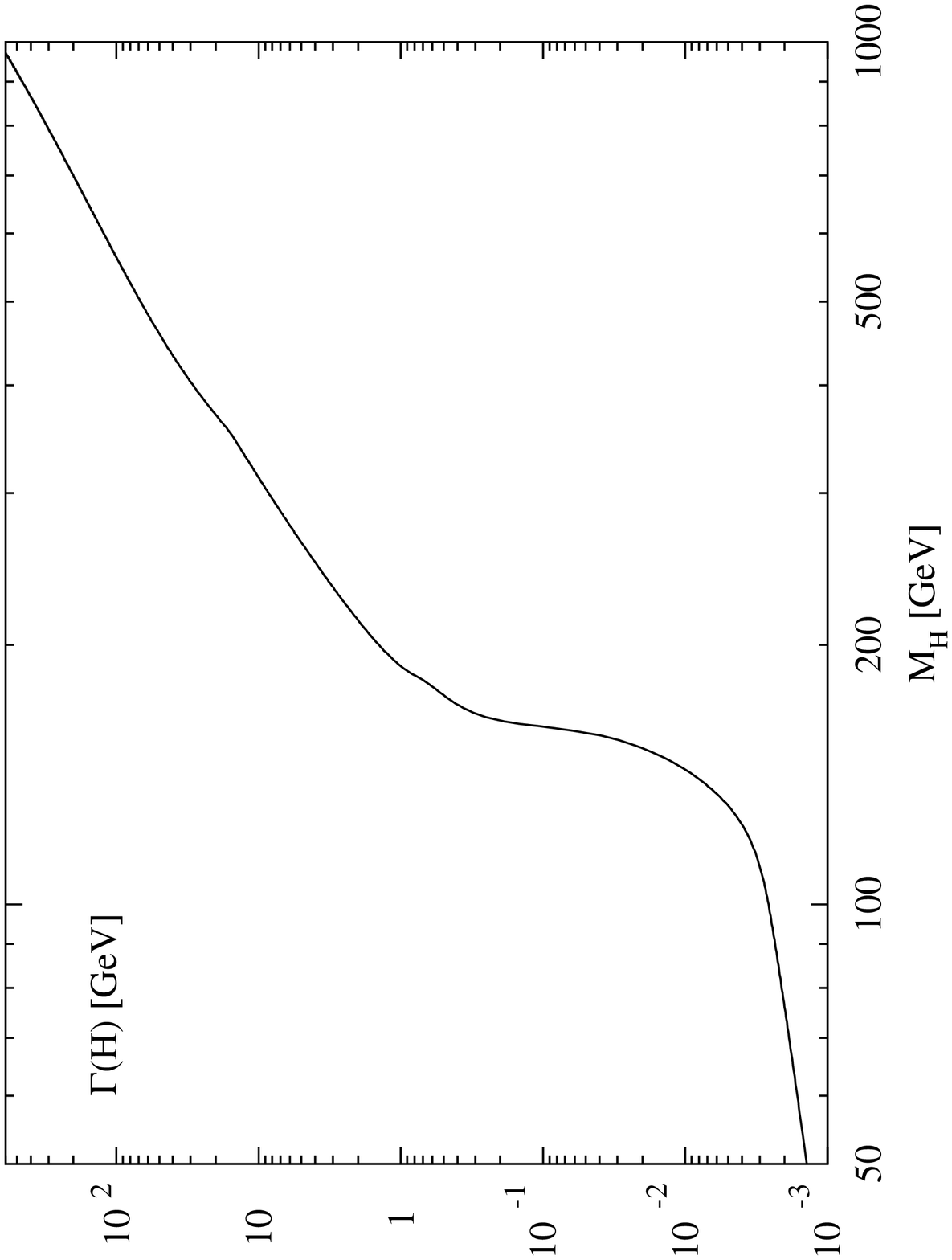}
\end{turn}
\vspace*{-5.5cm}

\hspace*{+6.5cm}
\begin{turn}{270}%
\epsfxsize=5.5cm \epsfbox{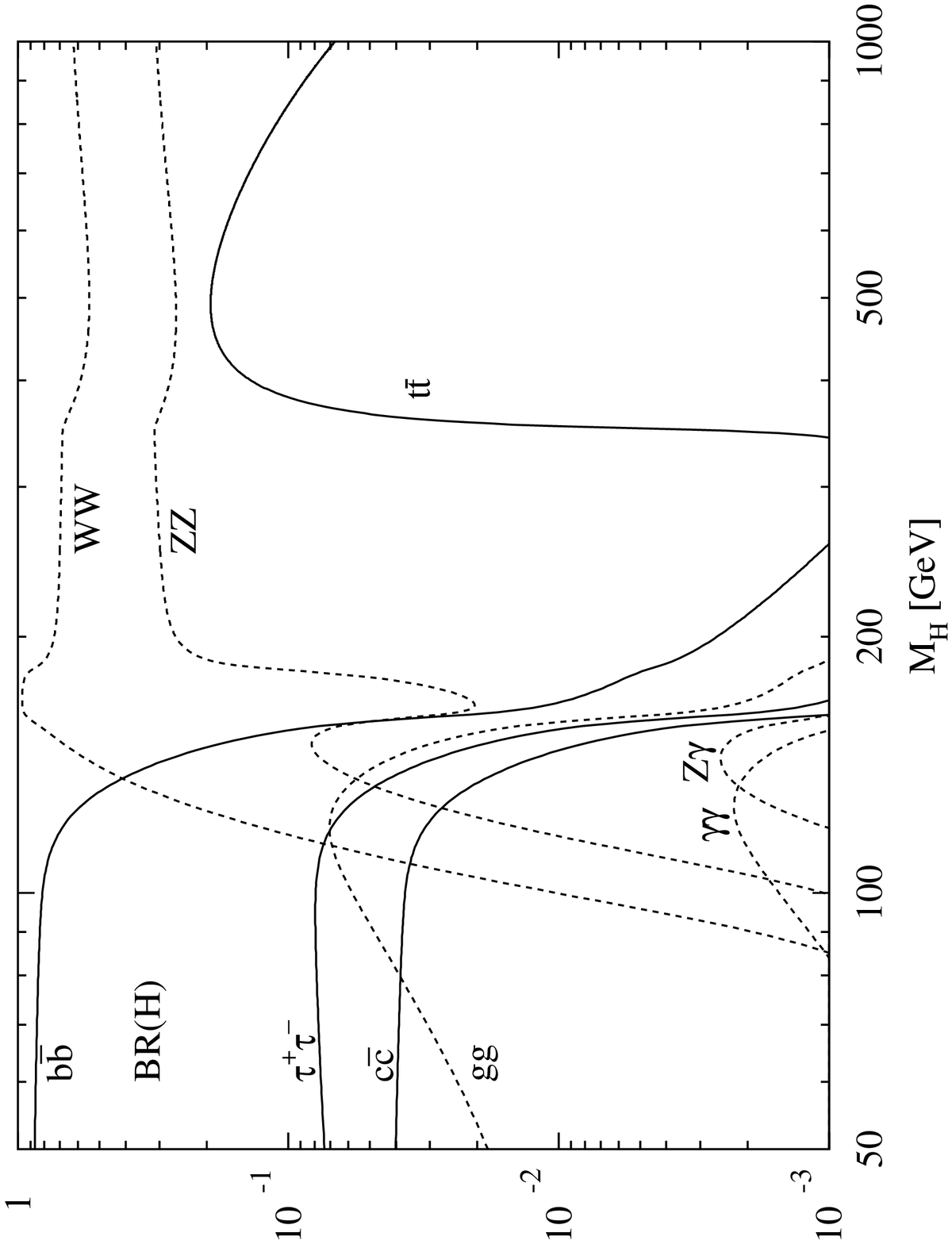}
\end{turn}
\vspace*{0.0cm}
\end{center}
\caption[]{ Total decay width $\Gamma(H)$ in GeV and the main decay modes
 $BR(H)$ of the Standard Model Higgs boson (from \cite{jan}).}
\label{fig:jan}
\end{figure}

Searches are based on the properties of the Higgs boson
expected within SM, like the  total decay width and the 
decay rates for the specific channels, see Fig.~\ref{fig:jan}.
It is obvious that the heavier the particle the larger coupling to 
 the Higgs particle is expected.

\begin{figure}
\center
\psfig{figure=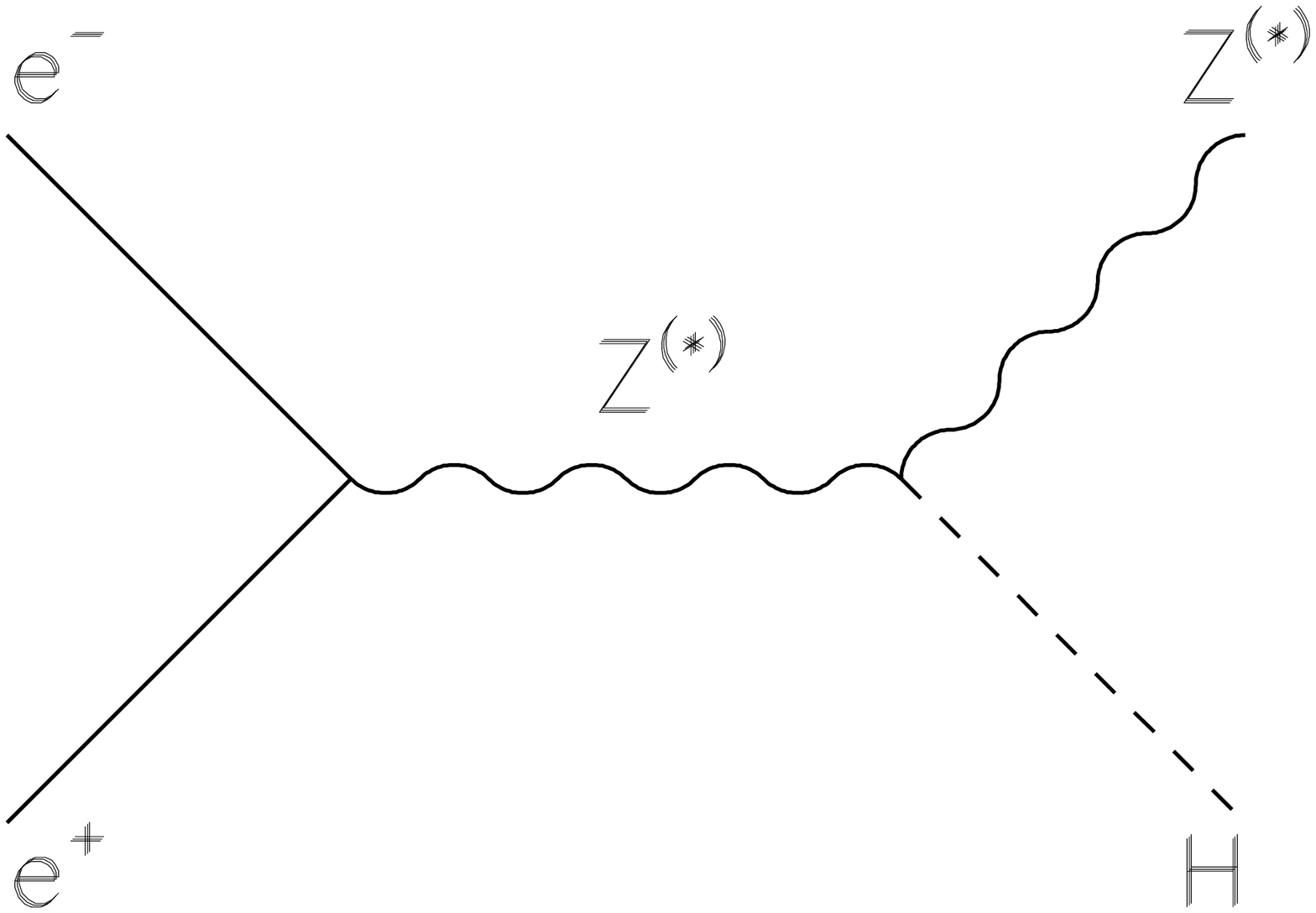,height=2.0in}
\vskip -2.0in
\hskip 0.0in
\psfig{figure=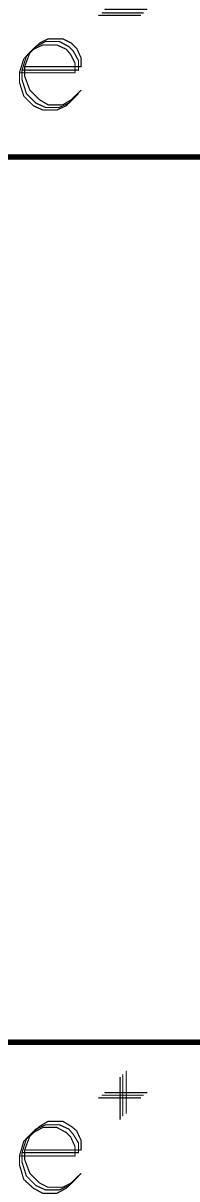,height=2.0in}
\caption{a) The Bjorken process; b) The fusion processes.}
\label{fig:diag}
\end{figure}

The most important constraint at LEP I (at the so called $Z-peak$,
\ie. at the CM energy $\sim M_Z$)
  arises therefore from the Bjorken process (Fig.~\ref{fig:diag}a)
\be
e^+e^-\ra Z^* \ra Z^* H_{SM}.
\ee
At LEP II also $W$'s and $Z$'s fusion processes start 
to constrain the mass 
of the SM Higgs particle (Fig.4b).

\begin{figure}
\center
\hskip 0.5in
\psfig{figure=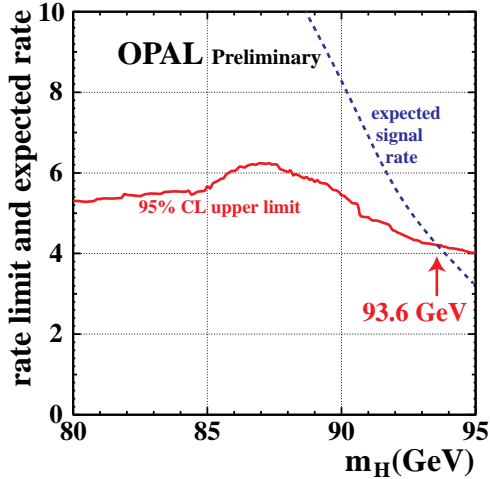,height=2.5in}

\caption{Results from the LEP II (energy 189 GeV) 
obtained by the OPAL Collaboration 
giving the highest limit for the SM Higgs scalar \cite{highest}.}
\label{fig:highest}
\end{figure}

The present combined 95 \% C.L. limit for the mass of the Higgs boson
from the direct search at LEP (performed up to the 
CM energy 183 GeV) is as follows 
\cite{limsm,desch}
\be
         M_{H_{SM}} > 89.8 \rm  ~~GeV.
\ee
In  Fig.~\ref{fig:highest} the highest mass limit (93.6 GeV)
 from LEP II (at the energy 189 GeV)
obtained by the OPAL group is  presented \cite{highest}. 

The indirect analysis, where  quantum effects due to the
various fundamental  particles of the SM are included, gives a hint that
the Higgs boson mass is low. The indirect analysis  
based on all precision data as collected and analyzed in 1998, gives 
 the SM Higgs mass (95\% C.L.) \cite{eww98}
\be
M_{H_{SM}}=66^{+74}_{{-39}} {\rm ~~GeV}.
\ee
One should stress a large sensitivity of the above limit to the input data. 

\begin{figure}
\begin{center}

\vskip -32mm
\hspace*{-1.0in}
\mbox{\psfig{file=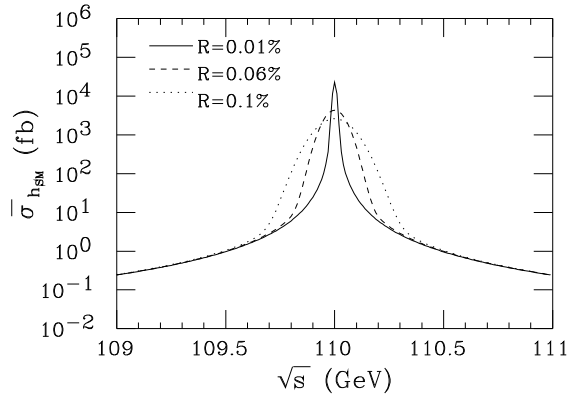,width=180mm}}
\vskip -150mm
\end{center}
\caption{ The Higgs boson  peak at Muon Collider, for three energy 
resolutions $R$.
From \cite{mugun}. 
\label{fig:mugun}}
\end{figure}

A unique opportunity to see a Higgs boson as a resonance
in the process $\mu^+\mu^-\ra h \ra f {\bar f}$ can appear 
at the Muon Collider \cite{mu}.  The expected results 
are presented in Fig.~\ref{fig:mugun}
 for mass of the SM Higgs boson  equal to  110 GeV \cite{mugun}. 
Note, that since the  Higgs boson width is expected to be  small
 (Fig.~\ref{fig:jan}a),
 the line shape will be given by the energy resolution. 

\section{Open problems of the Standard Model}
Although the Standard Model nearly perfectly, 
below two  standard deviations, describes the existing data
 there are still open problems 
which are not satisfactorily solved in the SM \cite{hollik}.
Large number of free parameters, 
the  lack of  explanation  for the existence of the generations of fermions, 
the  lack of  explanation of the mass pattern for fermions
(scales and mixing among them) can not satisfy particle physicists.

As follows from a discussion in Sec. 5 the indirect  Higgs boson mass
 bound  (32)
 means that the SM may be extrapolated up to energies 
around the Planck scale $M_{Pl}=10^{19}$ GeV,
where the quantum-gravity effects are expected to appear \cite{hollik,vac}. 
But the smallness of the electroweak scale $v\ll M_{Pl}$ 
seems then to be a problem (a ``hierarchy'' problem), 
which is related to the fact that quantum corrections to $v^2$
will contain term $\sim M_{Pl}^2$, strongly 
modifying the mass of the gauge boson
and destroying the hierarchy of scales.


The well known quadratic divergence, which appears in the Higgs boson mass
 squared due to the SM particle loop corrections, is another face 
of this problem. 
 Thus, it is not ``natural'' to have a
 light Higgs boson  in SM. The replacement of an elementary Higgs boson
 by a bound state might  cure  this problem. More promising seems to be
solution  based on  supersymmetry where  additional particles
(superpartners)  contribute in a way, which guarantees a cancellation 
of the divergencies.
Note, that ``naturalness'' leads to a   
scale $\cal O$ 1 TeV for mass of superparticles.  

The unification of gauge coupling constants cannot be obtained in the SM,
 pointing once more to a need of  larger  framework, \eg. supersymmetry.

{\sl It is a curiosity of the SM that (some) of these questions will 
persist even after the Higgs boson will have been discovered 
\cite{hollik}}.

A recent experimental evidence  for the massive neutrinos 
leads to  the additional questions - note that 
although  the degrees of freedom of the Standard Model 
permit neutrino masses, 
a larger theoretical context is needed  in order to understand it,
 see \cite{hollik,zralek}.

\section{Beyond the Standard Model}
No doubt, the most important ideas leading beyond SM 
are related to  supersymmetry.
The minimal version of such  model 
is called the Minimal Supersymmetric Standard Model (MSSM). 
 {\sl MSSM, mainly theoretically advocated, 
is competitive to the SM in describing the data with about the same 
quality in global fits
\cite{hollik}} (see also \cite{sp}). 

The unification of supersymmetric gauge theories 
with quantum gravity  within a superstring approach looks also 
appealing.
A unification of gauge couplings is a
 feature intrinsic to the theory.
Recently it has been speculated that the 
characteristic quantum-gravity scale could be  as low as the weak scale 
 \cite{giudice}.
If this is the case, one loses the original motivation for 
supersymmetry, based on the hierarchy problem.
{\sl Supersymmetry may still be desirable as a necessary ingredient of
 string theory, but it could be broken at the string level and not be 
present in the effective low-energy field theory {\cite{giudice}}.}

Other suggestions are going  
towards  the possibility of a non-supersymmetric 
1 TeV Grand Unification Theories \cite{dienes}.
It seems that there are good
reasons to discuss   extensions of the SM 
without supersymmetry.  Interestingly enough
the minimal non-supersymmetric model  Two Higgs Doublet Model II
(called here 2HDM) can also properly describe
the low energy data  even with one very light neutral
Higgs boson \cite{grant,ckz}.
   
\section{Two Higgs Doublet Extensions of SM}
One of the  minimal extension of the Standard Model is the approach where
instead of one, two complex scalar doublets of SU(2) (with $Y_W=1$) 
are introduced \cite{hunter}\\
$$
\begin{array}{cc}
 \Phi_1= \left( \begin{array}{c}
\phi_1^+ \\ \phi_1^0
\end{array} \right) &
 \Phi_2=\left( \begin{array}{c}
\phi_2^+ \\ \phi_2^0
\end{array} \right). 
\end{array}
$$
The potential is usually parametrized as \cite{hunter}
\bea
V( \Phi_1, \Phi_2)=\lambda_1( \Phi_1^{\dag}\Phi_1-v_1^2)^2
+                  \lambda_2( \Phi_2^{\dag}\Phi_2-v_2^2)^2\\ \nonumber
+\lambda_3[(\Phi_1^{\dag}\Phi_1-v_1^2)+( \Phi_2^{\dag}\Phi_2-v_2^2)]^2
\\ \nonumber
+\lambda_4[( \Phi_1^{\dag}\Phi_1) (\Phi_2^{\dag}\Phi_2)-
( \Phi_1^{\dag}\Phi_2) (\Phi_1^{\dag}\Phi_2)]\\ \nonumber
+\lambda_5[Re(  \Phi_1^{\dag}\Phi_2)-v_1v_2 \cos\xi]^2\\ \nonumber 
+\lambda_6[Re(  \Phi_1^{\dag}\Phi_2)-v_1v_2 \sin\xi]^2,
\eea
where the parameter $\xi=0$ guarantees a CP conservation. Below only this
 case will be discussed.

After spontaneous symmetry breaking two vacuum expectation values appear\\
$$
\begin{array}{cc}
 < \Phi_1>= \left( \begin{array}{c}
0 \\ {{v_1}\over{\sqrt2}}
\end{array} \right)  &
 < \Phi_2>=\left( \begin{array}{c}
0 \\ {{v_2}\over{\sqrt2}}
\end{array} \right) 
\end{array},
$$
with 
\be
v^2=v_1^2+v_2^2, {\rm ~~and~~} {M_W={{g v}\over{2}}}.
\ee
In the model with two (in fact any) scalar doublets (and singlets)
large corrections to the $\rho$ parameter are naturally avoided.
There are few patterns how  the new fields may couple to fermions 
in order to avoid in addition the flavour
 changing neutral currents (FCNC). 
Four models are considered in the literature (see \eg. \cite{hunter,santos}):  
in Model I only $\Phi_2$ couples to all fermions,
in Model II - $\Phi_2$ couples to the up quarks and  $\Phi_1$ to the down 
quarks and the charged leptons,
Model III - $\Phi_2$ couples to the up quarks and the charged leptons
 and  $\Phi_1$ to the down quarks, finally 
Model IV - $\Phi_2$ couples to the quarks and $\Phi_1$ to the leptons.
The Higgs sector of the  MSSM has a structure of  Model II,
therefore below I will concentrate only on this model.
First the common features of and limits 
for the Higgs  sector in the Model II will be presented
(Sec.8.1 and 8.2). In Sec.9 the results 
specific for the supersymmetric
version of the Model II, \ie. MSSM,   will be given, while in 
Sec.  10 the non-supersymmetric approach, 2HDM, will be discussed. 

\subsection{Model II}
 This model may
 explain a large ratio between the  top quark $t$
and the bottom quark $b$ mass by relating it to the 
large ratio of the vacuum
 expectation values \cite{large}:
\be
\tan \beta ={{v_2}\over{v_1}}.
\ee
The neutral Higgs fields 
may  mix among themselves, this  mixing is parametrized by the 
parameter $\alpha$.
The content of the Higgs sector of the Model II 
in terms of physical parameters can be described
 as follows (for the  CP conservation) 
$$ M_h,~~M_H, ~~M_A, ~~M_{H^{\pm}} {~~\rm and} 
~~\tan \beta, ~\alpha,~\lambda_5,$$
where $h,~~H$ are the neutral scalars (CP-even particles, by definition 
$M_h\le M_H$), $A$ is a CP-odd particle (often
called a pseudoscalar) and  $H^{\pm}$ denotes charged Higgs scalars.

Parameters $\tan \beta$ and $~\alpha$  govern the corresponding 
couplings of Higgs bosons to 
 themselves and to gauge bosons and fermions. 
The SM couplings of the $h$ to fermions,
 $(-igm_f/2M_W)$ are modified by multiplicative 
factors which differ for the two fermion isospins. 
 For example for bottom and top quarks we get: 
\bea
hb {\bar b}:{\hspace{0.4cm}} {{-\sin\alpha}\over{\cos\beta}}=
\sin(\beta-\alpha)-\tb \cos(\beta-\alpha)\\
ht {\bar t} :{\hspace{0.5cm}} {{\cos\alpha}\over{\sin\beta}}=
\sin(\beta-\alpha)
+{{1}\over{\tb}} 
\cos(\beta-\alpha)
\eea
The $h$ couples to $ZZ$ with the  SM factor
 ($ig {{M_Z} / {\cos\theta_W}} ~g^{\mu \nu}$) times
\be
hZZ: {\hspace {0.5cm}} \sin(\beta-\alpha). 
\ee
One can see that for $ \sin(\beta-\alpha)$ =1 the Higgs boson $h$  in
the general Model II behaves as the Higgs boson 
in the SM, while for  small $ \sin(\beta-\alpha)$ a large
differences between two approaches may appear. 

For the coupling of the pseudoscalar $A$ to fermions, 
the corresponding factors, with the normalization as for $h$ in the SM , are 
\be
Ab {\bar b} : {\hspace{0.3cm}} -i\gamma_5\tb;~~
At {\bar t} : {\hspace{0.5cm}} -i\gamma_5{{1}\over{\tb}}.
\ee 
The $AZZ$, $AWW$ couplings are absent in the considered model,
as a results of CP conservation \cite{hunter}.

For large $\tan \beta$  couplings of $h$ and $A$ to the down
 quarks and the charged leptons are strongly  enhanced while those 
 to the up  quarks are suppressed. 
The situation is reversed for low $\tan \beta$.

From a requirement of the perturbativity of the calculation 
 the value of $\tan \beta$
is restricted to lay  between  0.2 and 200 \cite{nir}. 

Below the constraints which are valid for both the supersymmetric 
and non-supersymmetric versions  of Model II will be  presented.
\subsection{Basic searches}
At LEP the basic Higgs boson searches in the neutral sector of the
 Model II are based on the following processes, 
\begin{itemize}
\item the Bjorken process $e^+e^-\ra Z^* h$,
\item  the pair production $e^+e^-\ra A h$,
\item the Yukawa process $e^+e^- \ra f {\bar f} h(A)$,
\end{itemize} 
where the  corresponding  Higgs couplings  
can be measured directly.

\begin{figure}
\begin{center}
\vspace*{-0.5in}
\hspace*{-2.5in}
\mbox{\psfig{file=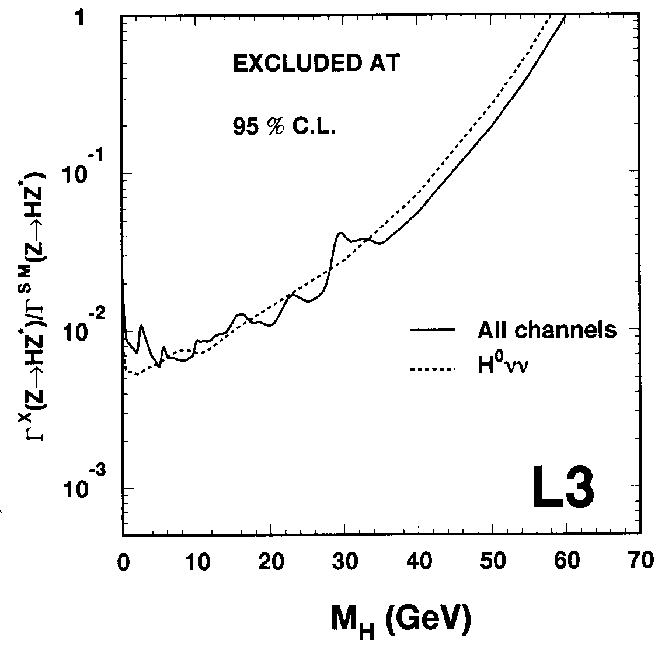,width=85mm}}
\hspace*{+2.5in}
\vspace*{-6.5in}
\mbox{\psfig{file=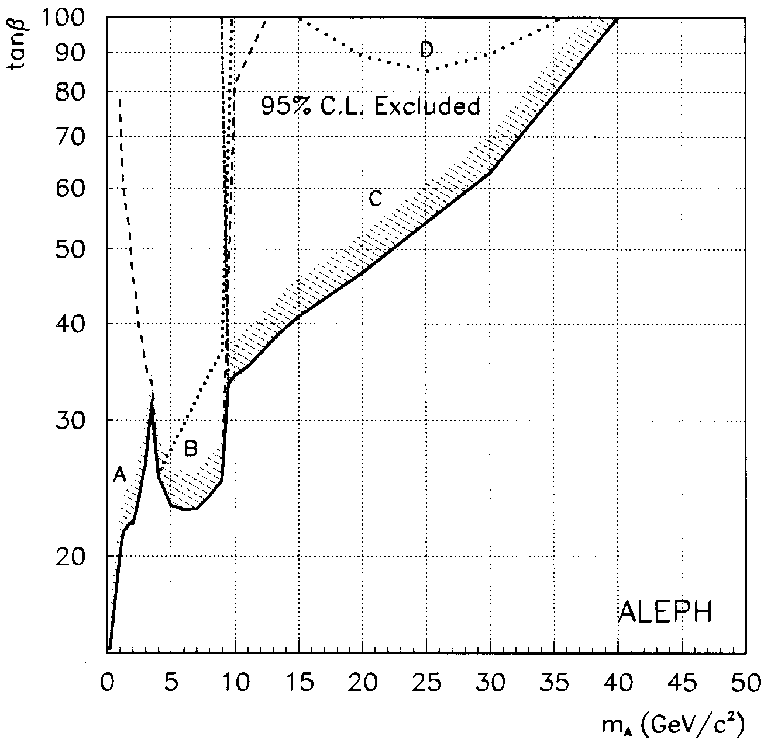,width=70mm}}
\end{center}
\vspace*{3.5in}
\caption{a) 
The  experimental limits on $\sin^2(\beta-\alpha)$
from L3 \cite{sinbal3}, b)
 the present limits for $\tan \beta$ versus mass $M_A$ 
from
 the Yukawa process $e^+e^- \ra {\bar f} f A $ at LEP I \cite{yukaleph}.
\label{fig:yuk}}
\end{figure}

The  cross section for the Bjorken process is given by 
$\sigma (e^+e^-\ra Z^* h)=\sin^2(\beta-\alpha) 
\sigma_{SM} (e^+e^-\ra Z^* h)$; therefore
by measuring it   one can constrain   the $h$ coupling to gauge boson
 (see Eq. 38 and Fig.~\ref{fig:yuk}a, and also Fig.~\ref{fig:highest}).
On the other hand the pair production cross section  sets a limit on the 
$\cos^2(\beta-\alpha)$.
By  combining  these two limits a allowed region of masses can be derived,
see below. 

The strength of the pseudoscalar coupling to the down-type  fermions, 
$\tan \beta$, 
 was studied at LEP I by the ALEPH group in the Yukawa process
$e^+e^-\ra Z \ra {\bar f} f A $ \cite{yukaleph}. Results are shown
in Fig.~\ref{fig:yuk}b.

For charged Higgs boson mass  there are constraints 
from the direct search performed at LEP for the process  $Z\ra H^+H^-$.
 The  95\% C.L. limit, deriven from  four LEP experiments,
 for  the CM energy up to 183 GeV, is \cite{desch}
$$M_{H^{\pm}}\ge 59 {\rm ~GeV}.$$

\begin{figure}[hbtp]
\begin{center}
\hspace*{-6.5cm}
\epsfxsize=5.5cm \epsfbox{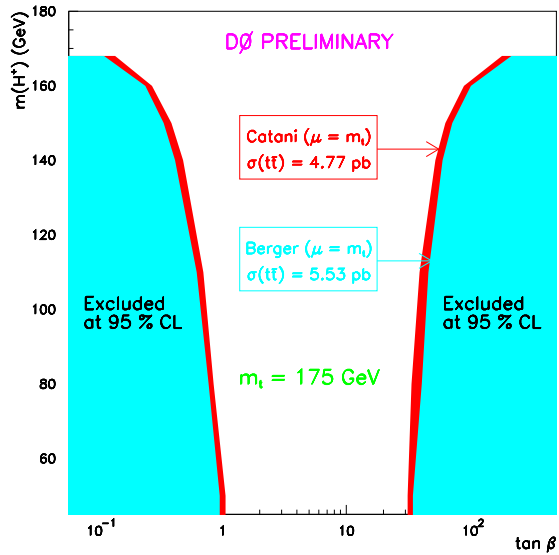}
\vspace*{0.0cm}
\end{center}
\caption[]{ The exclusion plot for the mass of the charged 
Higgs boson as function of $\tb$ \cite{TEV}.}
\label{fig:char}
\end{figure}

If the charged Higgs boson is lighter than 170 GeV,
the $t$ quark may decay into charged Higgs boson and $b$ quark. 
In such a case 
in addition to the LEP limit,  the limit for   $M_{H^{\pm}}$ 
as a function
of $\tb$ can be obtained  from the TEVATRON data\cite{TEV}.
The resulting exclusion region  in the mass $M_{H^{\pm}}$ 
and $\tb$ plane is presented in  Fig.~\ref{fig:char}.
Note, however that the EW and QCD (strong interaction) 
 corrections may invalidate  this   analysis in Model II, as it was 
pointed our recently \cite{sola}.

The process $Z\ra h(A) \gamma$, proceeding  via  loop diagram,
 may be of an importance
as it has been shown in a  recent study \cite{mkz}, see below Sec.10.
There is a sensitivity here not only to the $h(A)$ couplings
 to fermions and
gauge boson $W$ (for $h$), but also to the trilinear coupling 
$hH^+H^-$ ($\lambda_5$) and the  mass of $H^{\pm}$. 

Below we shortly present specific  constraints of 
the two versions of the Model II, the supersymmetric 
MSSM and non-supersymmetric 2HDM.
We will see how  much 
phenomenological consequences differ in these two approaches.
The most striking difference is that
much lighter Higgs bosons are allowed by the same data in the 2HDM 
than in the MSSM case.

There are many excellent reviews on the status of  Higgs particle searches in
the MSSM model (see \eg. \cite{kane}), therefore below only 
the main results
obtained in this model will be collected. In the following section (Sec. 10)
non-supersymmetric version of the Model II, \ie.  2HDM, will be disccused
in detail. 

\section{MSSM}
Supersymmetry  relaties fermions  and bosons
in such a way that for each fermion there exists ``its'' boson and vice versa
 (supersymmetric partners).
The supersymmetry has to be broken - otherwise  exactly equal masses
 of  both kinds of particles are  
expected, contrary to the observed spectrum of particles.

 After  the gauge symmetry is broken spontaneously
 the MSSM Higgs sector arises  in a form  of the discussed above Model II
 with five Higgs bosons. 
 In the MSSM there appear supersymmetric relations between parameters 
of the model, leaving only two of them as  
independent parameters at the tree level, \eg. $M_A$ and $\tb$.
 Therefore  tight constraints
from the data  on the Higgs boson masses are expected (with a 
surprisingly weak dependence on the other sectors
 of supersymmetric particles). 
Below we list the present constraints on the Higgs sector in MSSM.

\subsection{Constraints  on the Higgs sector 
in the MSSM}

In supersymmetry the strength of the Higgs self
 coupling is related to the gauge coupling, leading to
the  bound on the mass of the lightest scalar $h$, namely
\be
M_h \le 135 ~\rm {GeV},
\ee
 for large $\tan \beta$ (\cite{hollik,carena}). 
\begin{figure}
\begin{center}
\hskip -3.5in
\mbox{\psfig{file=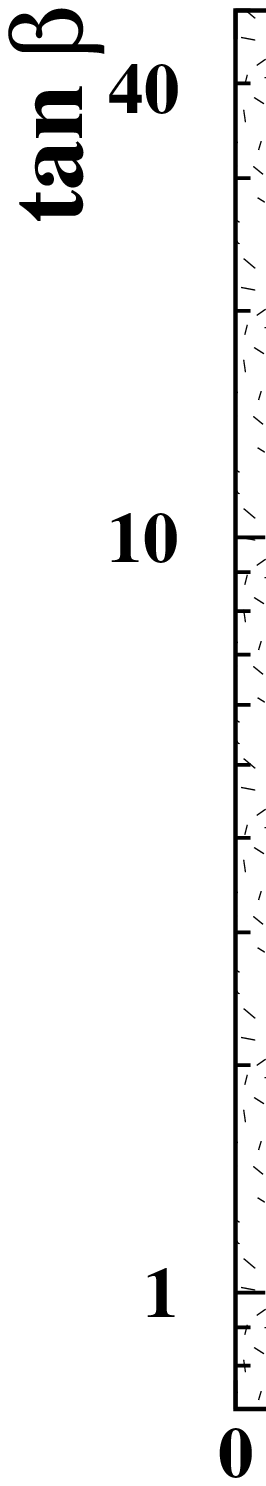,width=79mm}}
\hskip -3.5in
\mbox{\psfig{file=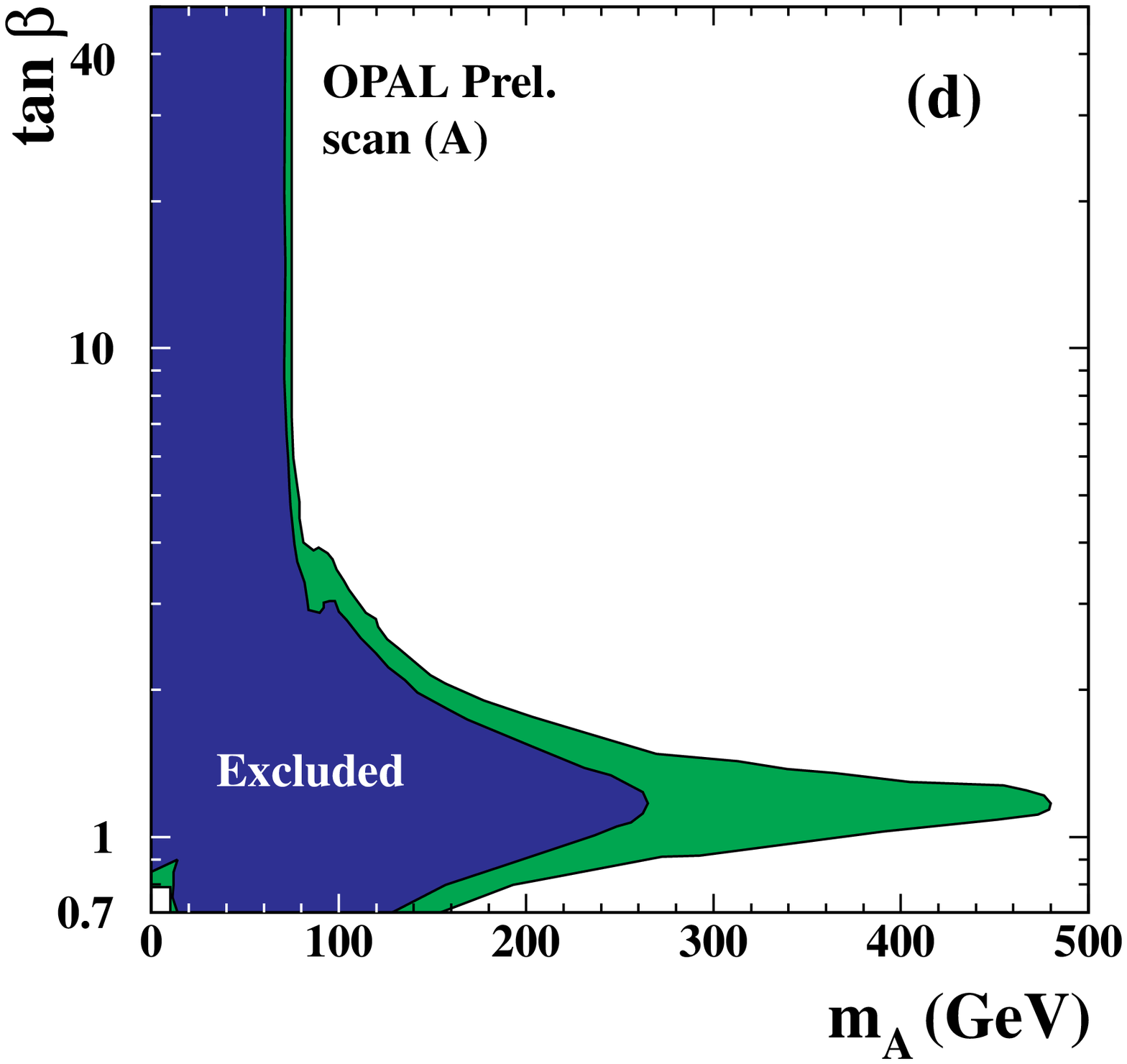,width=79mm}}
\end{center}
\caption{
 The OPAL  limits based on the LEP data at the CM energy 183 GeV
for a) $\tan \beta$ versus mass $M_A$, b)
$\tan \beta$ versus mass $M_h$
\cite{highest}. 
\label{fig:opal-mssm}}
\end{figure}

From the direct search  discussed above  (Bjorken process, 
and the pair production at LEP, see  Sec.8.2)
the analysis in the MSSM framework leads to 
the present  95\% C.L. limits  
(for the CM energies up to 183 GeV)
 \cite{desch,limsm}
$$    M_h\ge 77 {\rm ~GeV~~and}~~M_A\ge 78 {\rm ~GeV~~}, $$
for 
$$\tan \beta \le 0.8 {\rm ~~or} ~~\tan \beta \ge 2.1.$$
In Fig.~\ref{fig:opal-mssm}a,b 
the OPAL limits  for the neutral pseudoscalar and scalar 
based on  LEP data (with the CM energy 183 GeV) are shown
for $\tan \beta$ as a function of $M_A$ and $M_h$.

\begin{figure}[hbtp]
\begin{center}
\hspace*{-6.5cm}
\begin{turn}{270}%
\epsfxsize=5.5cm \epsfbox{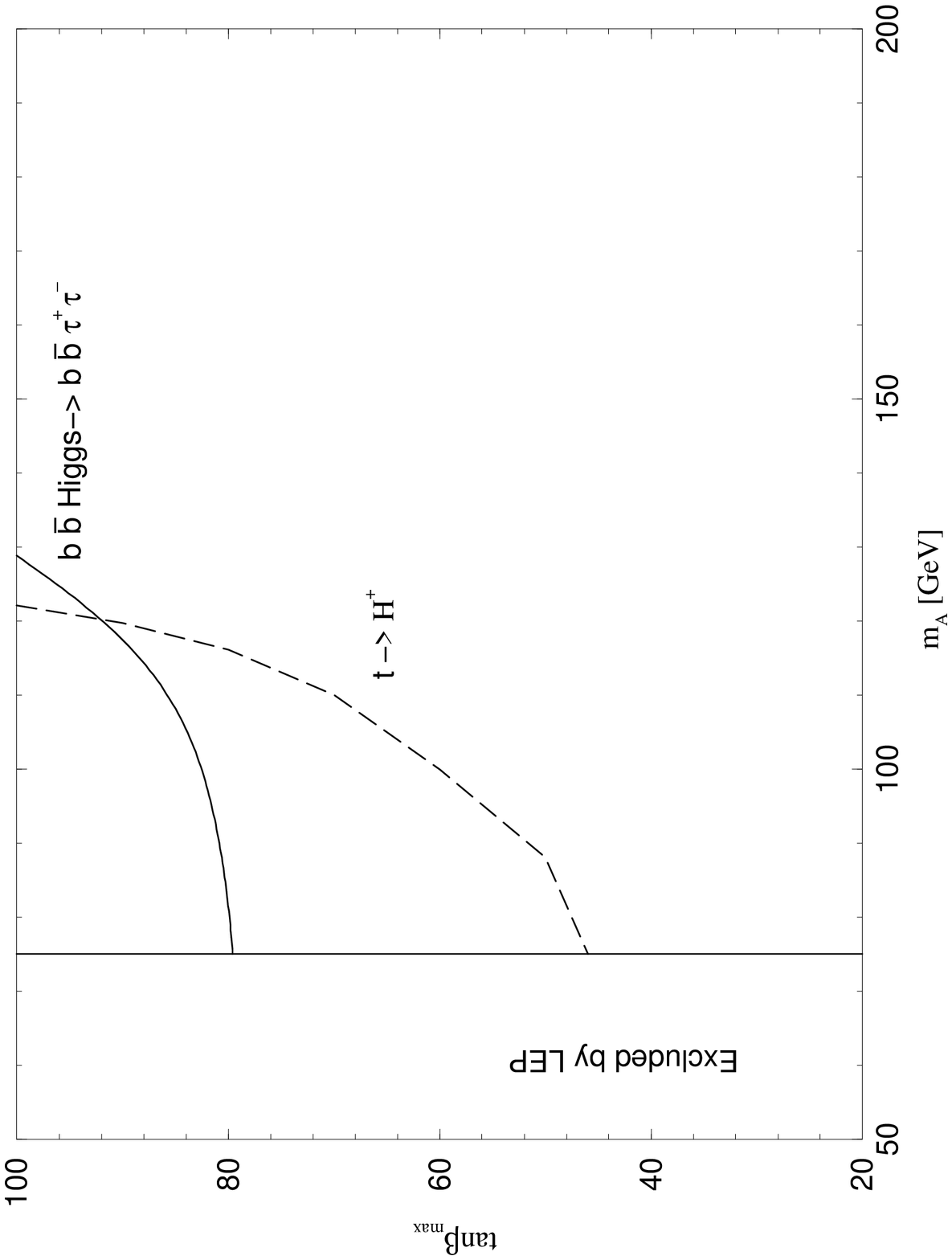}
\end{turn}
\vspace*{-10.0cm}

\hspace*{+6.5cm}
\epsfxsize=10.5cm \epsfbox{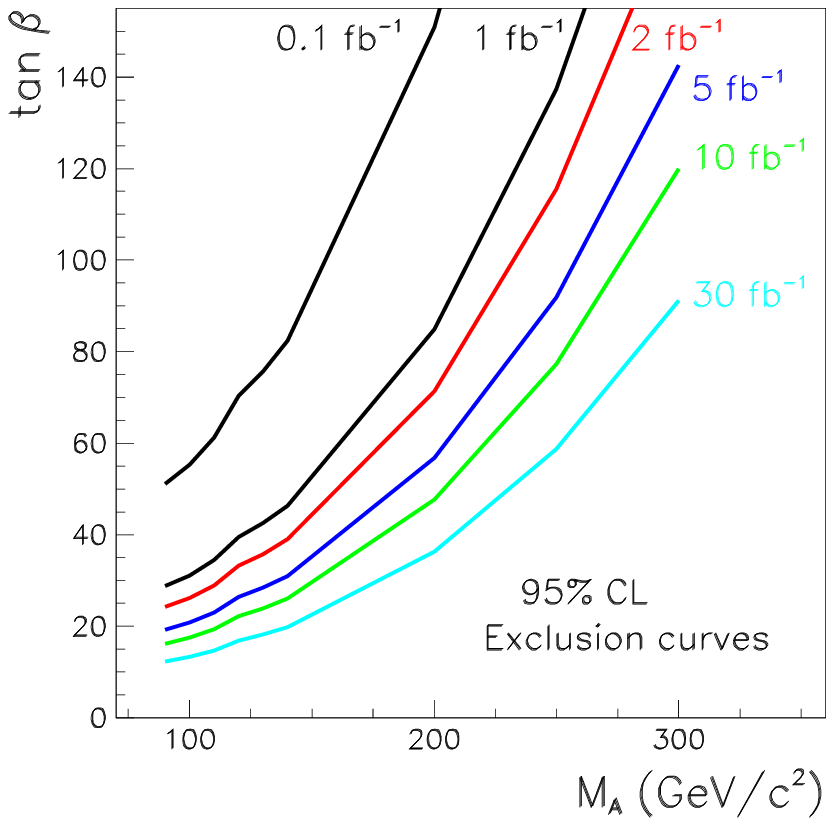}
\vspace*{0.0cm}
\end{center}
\caption[]{a) The present 95\% C.L. limits for $\tan \beta$ as a function of
$M_A$ in MSSM from the TEVATRON (I) data \cite{drees}.
b) Similar limits (based on $p {\bar p}\ra b {\bar b} A$) 
expected for the TEVATRON (II) for different  luminosities \cite{roco}.}
\label{fig:teva}
\end{figure}

The TEVATRON allows to constrain  the neutral Higgs sector as well. 
The maximal allowed $\tb$ as a function of the $M_A$ 
 from present data (Run I), and 
 expected for Run II are presented in Fig.~\ref{fig:teva}a and b,
 respectively.
(See also \cite{marcela}).

\begin{figure}
\begin{center}
\hskip -0.7in
\mbox{\psfig{file=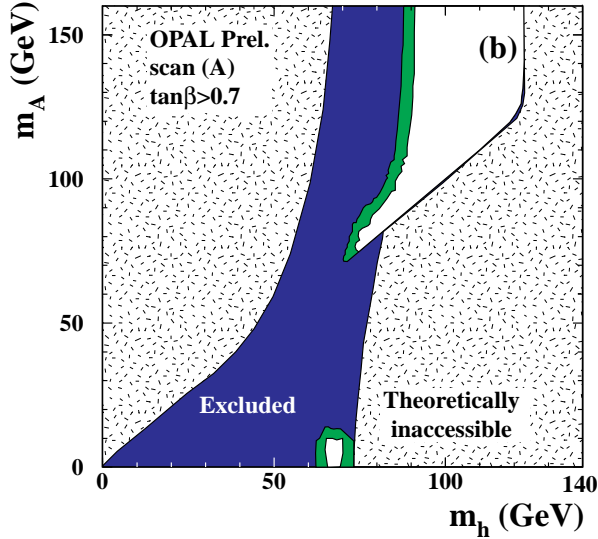,width=79mm}}
\end{center}
\caption{a) The  limits for  Higgs boson masses  $M_A$ versus  $M_h$ from OPAL
analysis obtained in  MSSM \cite{highest}. 
\label{fig:mham}}
\end{figure}

The allowed mass region for neutral Higgs bosons 
($M_h,M_A$) in the MSSM is shown in  see Fig.~\ref{fig:mham}. 
The important mass relation exists in the MSSM between pseudoscalar  and 
charged Higgs bosons.
The charged  Higgs boson mass has to satisfy the relation (at the tree level): 
\be
M_{H^{\pm}}^2=M_W^2+M_A^2.
\ee
Therefore we conclude that  both neutral and charged Higgs bosons with
 masses below, roughly, 80 GeV are excluded in the MSSM.

\begin{figure}
\begin{center}
\vskip -40mm
\hspace*{-1.0in}
\mbox{\psfig{file=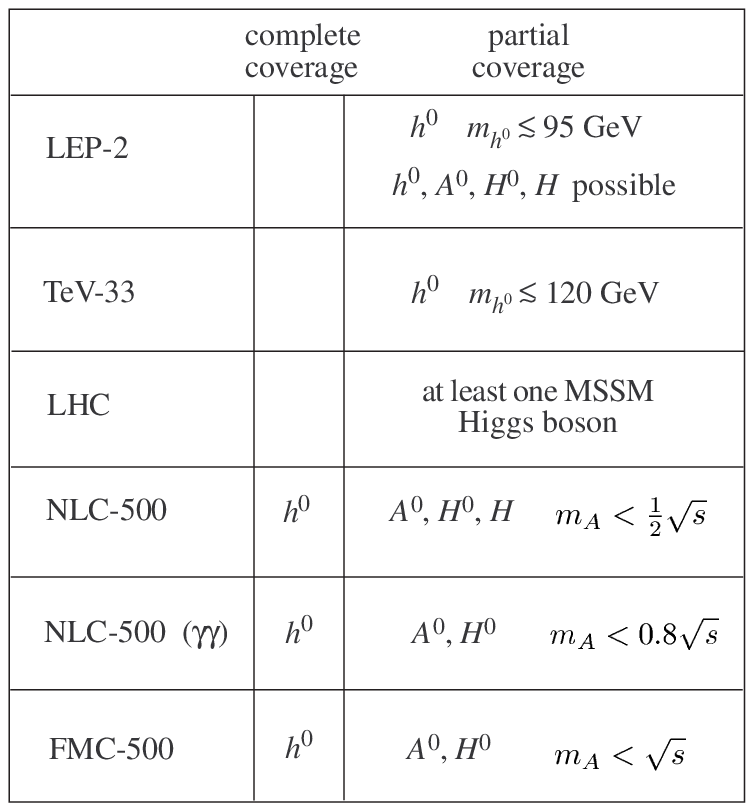,width=180mm}}
\vskip -235mm
\hspace*{1.8in}
\mbox{\psfig{file=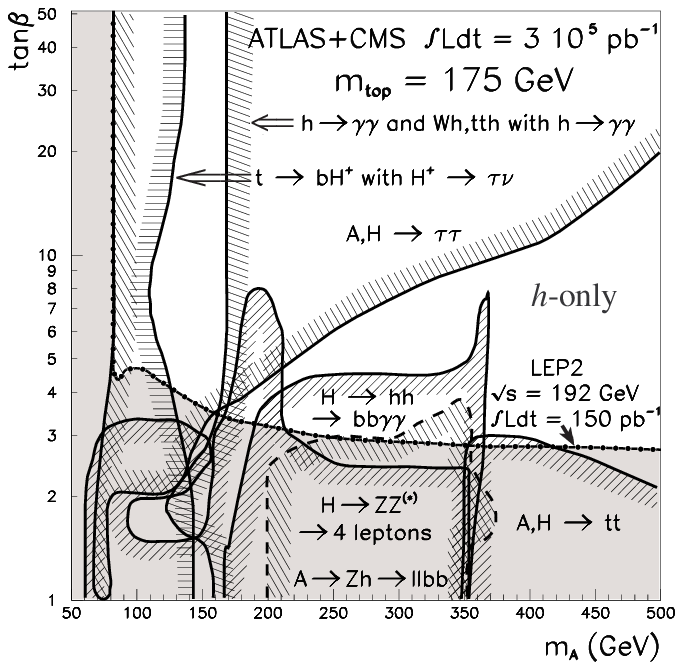,width=225mm}}
\vskip -170mm
\end{center}
\caption{a) The coverage of the parameter space in the MSSM by 
future colliders:
LEP II, TEVATRON (Run II), Next Linear Colliders at CM energy 500 GeV
(also with the $\gamma \gamma$ collider option) and the First Muon Collider
at the same enrgy   \cite{barger},
b) the potential of LHC for testing MSSM \cite{was}. 
\label{fig:cov}}
\end{figure}

The coverage of future searches of the Higgs bosons in the MSSM is summarized
 in Fig.~\ref{fig:cov}a, where the potential of different experiments 
is displayed and in Fig.~\ref{fig:cov}b, where the role of 
different channels
 at LHC is  shown  \cite{barger,was}.

To summarize:
{\sl If $M_h$ exceeds 130 GeV \footnote{in \cite{hollik,carena} the limit 
 135 GeV is given}, 
the MSSM is inconsistent.  
If $M_h$ is below 130 GeV, the MSSM is viable, but is the Higgs boson 
the MSSM Higgs?
Discovery of only one Higgs boson is insufficient to establish the MSSM
\cite{barger}.}

\section{2HDM}
The non-supersymmetric version of the Model II, 2HDM, 
 has a  Higgs sector the same 
as  MSSM but the relations among parameters imposed by the supersymmetry 
are missing. \footnote{Also the additional particles  besides 
 the Higgs bosons are not considered.} 
In contrast to the MSSM,   each parameter has to be constrained independently.
 
The requirements of vacuum stability and validity of perturbation theory
   suggest that 2HDM can not be valid up
 to the unification scale \cite{nie}. This is exactly   
what is expected if 2HDM
  is treated as a low energy realization of some more fundamental theory. 
\subsection{Search for  Higgs bosons in 2HDM}
The basic searches
are being  performed at LEP (see Sec. 8.2).
In addition some limits can be derived
 from the present measurement of the anomalous magnetic moment  
 for muon $(g-2)_{\mu}$, see
  \cite{g2}a,c \footnote{similar analysis
in the MSSM were performed as well}.

\begin{figure}
\begin{center}

\vspace*{0.5in}
\hspace*{-1.3in}
\mbox{\psfig{file=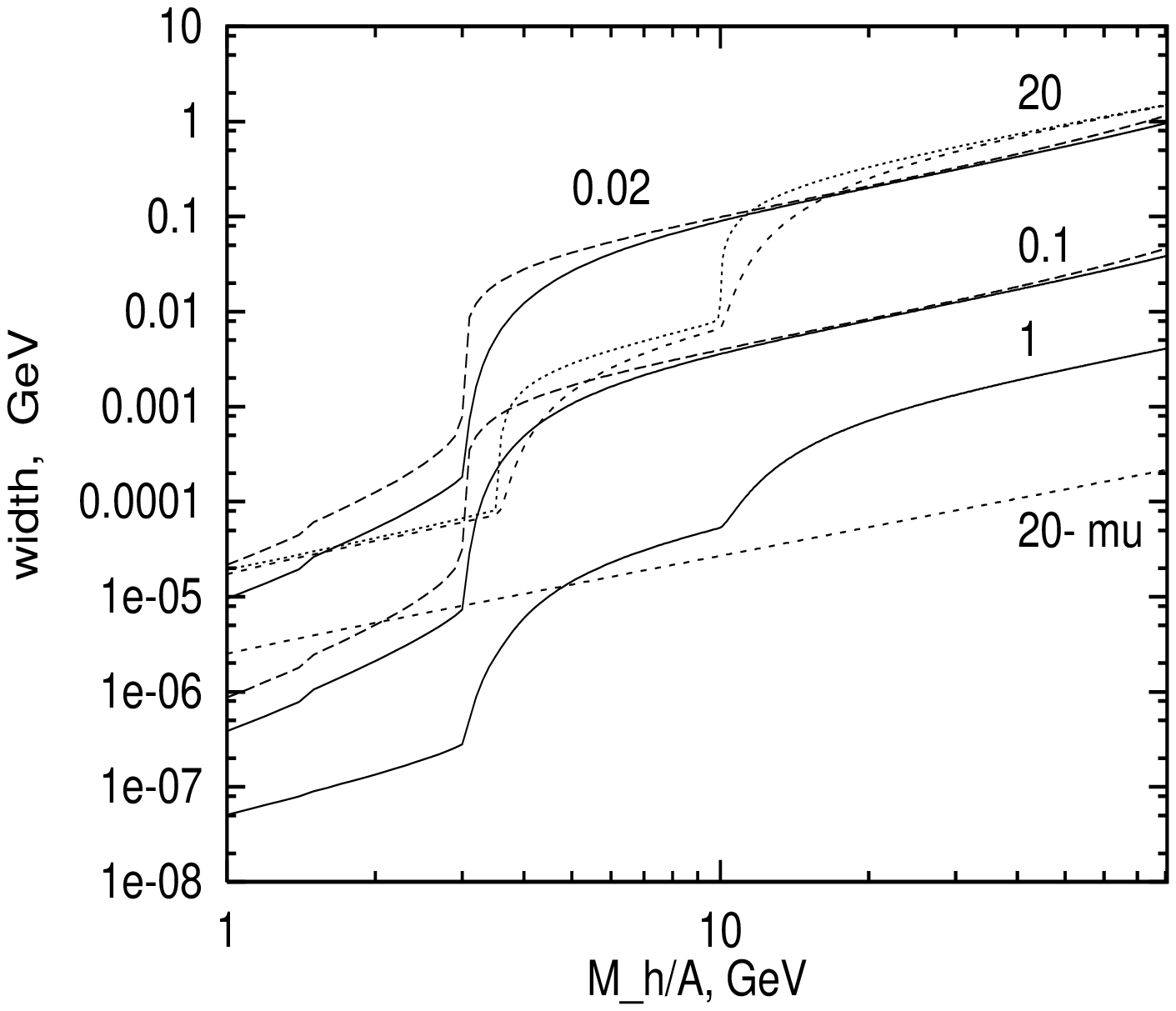,width=79mm}}
\hspace*{-0.1in}
\mbox{\psfig{file=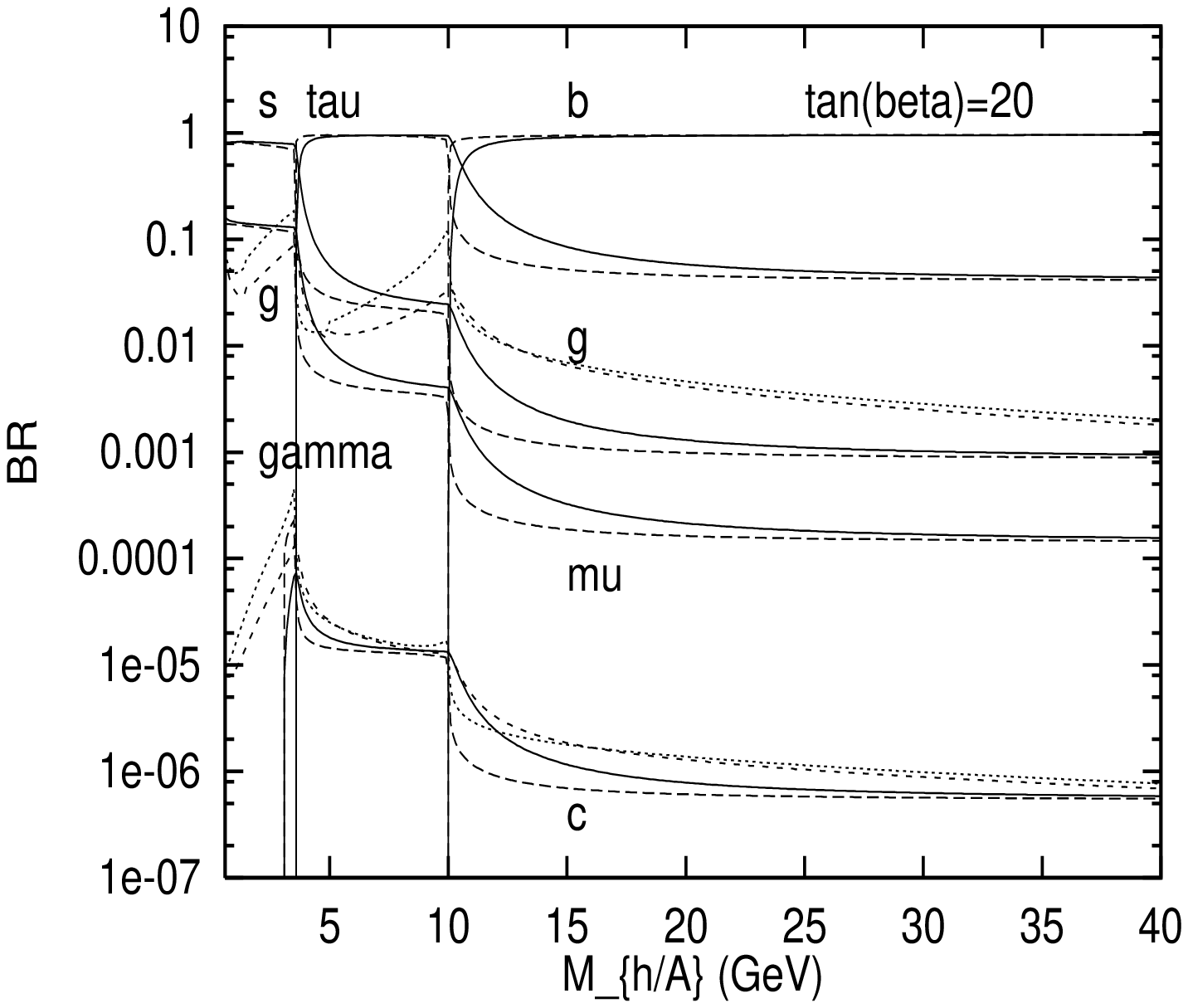,width=79mm}}
\end{center}
\caption{
a) The total width for $h$ and $A$ for $\tb$=0.1,~1 and 20 in the 2HDM.
For comparison results for  $\tb$=0.02, and the muonic partial decay width 
 for $\tb$=20
are shown \cite{mkk}, b) The branching ratio for $\tb$=20 is presented
 for $A$ and $h$ (with $\beta=\alpha$) \cite{mkz}.
\label{fig:bra2}}
\end{figure}

Direct searches 
 are  based on the expected  Higgs boson decay
 width and the branching ratios,
as shown  in Fig.~\ref{fig:bra2} for low and large $\tb$.  

\begin{figure}
\begin{center}
\mbox{\psfig{file=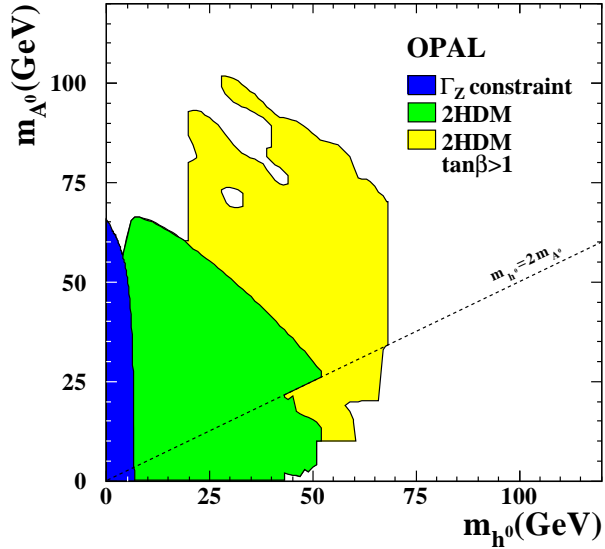,width=79mm}}
\end{center}
\caption{ The  \% 95. C.L. exclusion region for  Higgs boson masses,
  $M_A$ versus  $M_h$, from OPAL
analysis obtained in  the 2HDM \cite{opal366}. 
\label{fig:mha}}
\end{figure}

Combining the data from the Bjorken process and the pair production
leads to the exclusion mass region presented in Fig.~\ref{fig:mha}.
This is to be
 compared to Fig.~\ref{fig:mham},
 where a similar   exclusion plot is  obtained in the MSSM framework.
 The main conclusion is that in the 2HDM 
 a neutral Higgs particle lighter than in SM and in MSSM is allowed,
provided the other Higgs particles are heavy enough, 
$M_h+M_A\ge 50 $ GeV. This is also in agreement  with  recent results
from the global fit to the precision EW data obtained in the 2HDM \cite{ckz}.

Two scenarios are worth to be studied here:\\

$\ra$ with a (very) light scalar $h$\\

$\ra$ with a (very) light pseudoscalar $A$.\\

Note that the limit on the coupling of  $A$
to  $b$ quark and leptons ($\mu$ and $\tau$)  is given by the
measurement of the Yukawa process, see Fig.~\ref{fig:yuk}b for results.
Even  a very light pseudoscalar can couple with a large strength,
$\tb \sim$ 10-20.

The direct search  of a charged Higgs boson leads to results
 as discussed above in Sec. 8.2.
The indirect  limit on the mass of a charged Higgs boson
arises from the process  $b\ra s \gamma$. 
This process is mediated by loops  and therefore it is a 
probe of the Standard Model and of its  possible extensions.
In the context of the 2HDM one gets, 
for $\tan \beta$ larger than 2, 
a bound \cite{greub}
$$M_{H^\pm} \ge 165 \rm ~GeV,$$ 
which is not valid in MSSM.
Note, that this limit still allows (in 2HDM!) for the decay of $t$ quark 
to the charged Higgs boson 
and $b$-quark.

\begin{figure}
\begin{center}
\mbox{\psfig{file=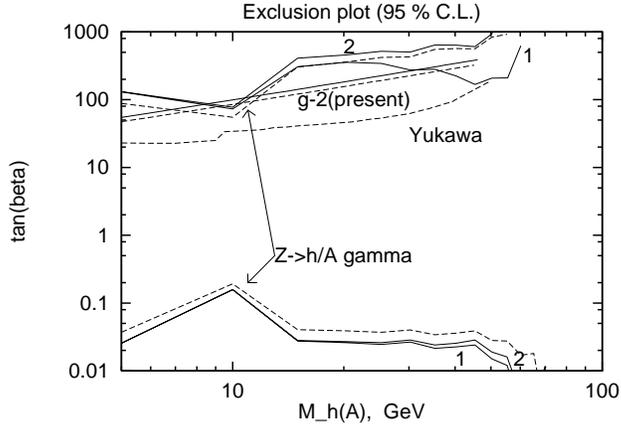,width=85mm}}
\end{center}
\caption{
 The present limits for $\tan \beta$ versus mass $M_h$ (solid lines)
 or $M_A$ (dashed lines) from
the analysis of  $Z\ra h(A) +\gamma$ process at LEP I,
compared to constraints from  the $g$-2 for muon data \cite{g2}
 and the Yukawa process $e^+e^- \ra {\bar f} f A $ at LEP I \cite{yukaleph}.
The regions above the upper and below the lower curves are excluded.
For the scalar production  experimental limits on $\sin(\beta-\alpha)$
from L3 \cite{sinbal3} are included and two masses of the
 charged Higgs boson are assumed: 1) 54.5 GeV and 2) 300 GeV. 
 From \cite{mkz}a.
\label{fig:zha1}}
\end{figure}

With the above  limits valid in 2HDM in mind
 we can discuss now new results from
the analysis of  the $Z\ra h(A) + \gamma$ process, measurements of which
were performed recently at the $Z$-peak by all four LEP experiments.
The measured branching ratio  is between  10$^{-6}$ and 10$^{-5}$
 \cite{brzhag,mkz}.
In the SM the scalar production  is due to the $W$
 and  fermion loop contributions 
(with a strong domination of the $W$-loop).
 The data lay above the  SM Higgs boson prediction.

In the 2HDM the decay $Z\ra h+\gamma$
 proceeds  via   loops with $W$ and fermions
(with    couplings
depending on the parameters $\alpha$ and $\beta$ (Sec.8.1)),
and in addition via  a charged Higgs boson loop. 
For the pseudoscalar production        
only fermion loops contribute.
The results are given Fig.~\ref{fig:zha1} in the form of  
the 95 \% C.L. exclusion plot
for $\tan \beta$ versus  $M_h$ or $M_A$.

\begin{figure}
\begin{center}
\mbox{\psfig{file=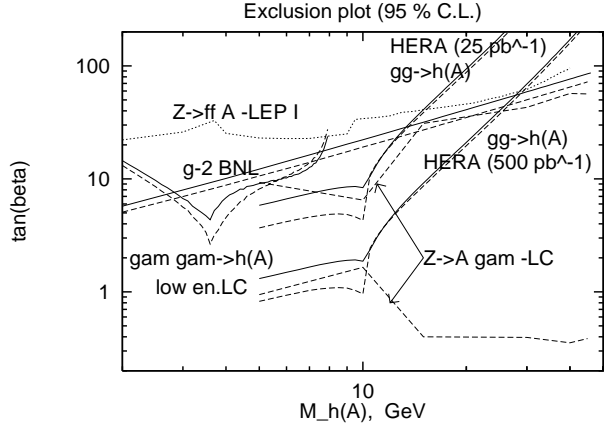,width=85mm}}
\end{center}
\caption{
 The potential of the future data on $g-2$ for muon 
\cite{g2}, the HERA measurement
(the integrated luminosity 25 and 500 $pb^{-1}$),
and the Linear Collider running at low energy $\sqrt s_{ee}$=10 GeV
(with 10$fb^{-1}$)
 and at $Z$-peak (with 20 $fb^{-1}$). For the reference the results 
from the Yukawa process $Z\ra {\bar f}f A$ at LEP I are shown.
The area above curves are excluded, with exception of the 
 $Z \ra A \gamma$, where area above upper and below lower curve are excluded.
From \cite{mkz}b. 
\label{fig:zha2}}
\end{figure}

An interesting opportunity to  look for a light 
neutral Higgs bosons  
is due to the photoproduction processes 
at the $ep$ HERA collider \cite{gpr}. 
Here the   Higgs boson production
 with masses below 40-50 GeV
is dominated by subprocesses 
due to the partonic content of the photon \cite{bk,mk}. 
In particular the process, where the gluonic content of the photon,
denoted as $g^{\gamma}$,
 interacts  with the gluon from the proton, $ g^{p}$,  producing a $h$ or $A$:
\be 
g^{\gamma} g^{p} \ra h (A),
\ee
with subsequent $h(A)$ decay into $\tau$ pairs was studied in detail \cite{bk}.
We found that for this channel 
one can, at least in principle,  get rid of a serious background
due to $\gamma g^p \ra \tau^+ \tau^-$. 
(For the    $b {\bar b}$ final state
the background is too large.)
The potential  of the HERA collider to search for a light Higgs boson
is larger than it follows from this analysis, since in addition there
are other subprocesses which contribute.
 On the other hand the SM Higgs search
is hopeless  at HERA due to the very small rate for  mass above 60 GeV
\cite{bk,mk}.

 The potential of futures searches of a very light neutral Higgs boson
 in the context of 2HDM is presented in Fig.~\ref{fig:zha2}.
Here the expected limits for the $\tan \beta$ versus Higgs boson mass 
from a new measurement on $g-2$ for muon at BNL
are presented together with a 
possible  exclusion 
based on the gluon-gluon fusion into $h$ or $A$ (Eq.~42) at HERA.
Results
 based on the neutral Higgs boson production in $\gamma \gamma$  
collision (with $h(A)$ decaying into muons) at low energy LC 
{\footnote{
 The very low energy $\gamma \gamma$ collider
has been suggested some time ago as a test machine for the NLC \cite{lowtest}.
The potential of such a  collider 
 in searching for a light neutral Higgs boson in 2HDM 
was studied in \cite{lowlc}.}},
 and 
the possible results from the process $Z\ra A \gamma$ 
measured at the LC running at the Z-peak are also given.   

\section{Future Colliders}
LEP is still collecting data, now at the CM energy 189 GeV.
The upgrading of the existing colliders TEVATRON and HERA
will allow to extend the direct searches of the Higgs boson(s) in the  SM, 
MSSM and 2HDM. Also a new high precision measurement
of $g-2$ for muon at BNL may provide crucial, although indirect,
source of information on the Higgs sector.   

The new hadronic collider LHC will cover a large part of  parameter space
of  the SM and MSSM. 
After the discovery of the Higgs boson, the determination of its 
properties will  be of highest priority. 
 New generation of the $e^+e^-$ Linear Colliders
with energy 300-500 GeV are considered to 
 be an ideal laboratory for discovering and
studying the intermediate-mass Higgs boson.

The Linear Colliders running as $\gamma e$ and $\gamma \gamma$
 Linear Colliders, with  high energy photon beams
offer excellent probe of the $Zh\gamma$, $ZA\gamma$ and 
$\gamma \gamma h $ or $\gamma \gamma A$ couplings.
All these  couplings  are of great importance for testing
the structure of the Standard Model and of the MSSM or 2HDM.
The Muon Collider can run at the Higgs boson - peak 
offering  a unique chance to 
study properties of the Higgs boson. 

\section{Conclusion and outlook}
No  Higgs boson has been  discovered so far.
Both in the SM and the  MSSM  there are hints that scalar Higgs $h$
 should be light, while in the 2HDM one neutral Higgs boson $h$ or $A$
may be light (even very light).  So, there is a good chance
to learn more about the origin of the
spontaneous symmetry breaking in the EW sector in  near future.

\vskip 0.5 cm
\noindent
I am grateful to the Organizers for the invitation of this excellent School. 
I am very much indebted  to P. Chankowski and 
J. Kalinowski for a critical reading of the
manuscript and   important suggestions,
and to K. Desch for sending the figure ~\ref{fig:mha}. 
Also a  help from J. Rosiek, U. Jezuita-D\c{a}browska and J. \.Zochowski
in preparing this contribution is acknowledged.


\begin{thebibliography}{99}
\bibitem{pdp98} C. Caso et al., {\em Eur. Phys. J.} {\bf C3} (1998) 1.
\bibitem{nambu} Y. Nambu, {\em A Matter of Symmetry}, 
The Sciences, May/June   (1992) 37.
\bibitem{sm} S. L. Glashow, \NP {\bf 20} (1961) 579; S.Weinberg \PRL {\bf 19}
 (1967) 1264; A. Salam, in Elementary Particle Theory,
 ed. N. Svartholm (Almqvist and Wiksells, Stockholm 1968.\\
H. Fritzsch and Gell-Mann, Proc. XVI Int. Conf. on High Energy Physics, eds. J.D. Jackson and A. Roberts, FERMILAB, 1972.
\bibitem{vanstein} A. I. Vainstein et al.,
{\em {Sov. Phys. Usp.}} {\bf 23} (1980) 429.
\bibitem{hunter} J.F. Gunion et al,{\em Higgs Hunter's Guide} 
(Addison-Wesley Publ. Comp., 1990);
\bibitem{higgs} P. W. Higgs, \PRL {\bf 12} (1964) 132, \PRep ~{\bf 145}
 (1966) 1156;
F. Englert and R. Brout, \PRL {\bf 13} (1964) 321; G. S. Guralnik et al., 
\PRL {\bf13} (1964) 585.
\bibitem{hollik} W. Hollik, Plenary Talk at the ICHEP'98, Vacouver
(hep-ph/9811313). 
\bibitem{vac}
T. Hambye, K. Riessekmann, \PR {\bf D55} (1997) 7255.
\bibitem{desch} K. Desch, Beyond SM Higgs Searches at LEP, talk in parallel 
session 10 at
ICHEP'98, Vancouver, Canada.
\bibitem{carzer} M. Carena and P. Zerwas, Higgs Working Group Physics
at LEP2, ed. by G. Altarelli, T. Sj\"{o}strand, F.Zwirner, 
CERN Report, N0. 96-01.
\bibitem{futtev} {\em Future EW Physics at TEAVATRON}, Fermilab-Pub-96/082,
eds. D. Auide, R. Brock (hep-ph/9602250).
\bibitem{lhc} V. Barger, hep-ph/9808354.
\bibitem{jan} A. Djouadi et al., {\em Comput.Phys.Commun.} {\bf 108} (1998) 56.
\bibitem{limsm}
LEP Higgs Working Group, ALEPH 98-069 PHYSIC 98-028,DELPHI 98-144 PHYS 790,
L3 Note 2310,
OPAL Technical Note TN-558, July 1998.
\bibitem{highest} OPAL Coll., PN-361, contribution to ICHEP'98.
\bibitem{eww98} LEP Higgs Working Group, LEPEWWG/98-01,
A combination of preliminary electroweak measurement 
and constrints on the standard model, May 98.
\bibitem{lc} E. Accomando et al., \PRep ~{\bf ~299} (1998) 1;
 V. Telnov, hep-ex/9810019;
R. Brinkmann et al., \NIM {\bf ~A406} (1998) 13.
\bibitem{mu} V. Barger et al., \PRep ~{\bf 286} (1997) 1.
\bibitem{mugun} J. F. Gunion, Physics at a Muon Collider, 
AIP Conference Proceedings 435, {\em Workshop on Physics at 
the First Muon Collider and at the Front End of the Muon Collider}, 
FERMILAB , Nov. 1997, p. 37.
\bibitem{zralek} M. Zra\l ek, \APP {\bf B29}, this proceedings (1998).
\bibitem{sp} P. Chankowski and S. Pokorski, \APP {\bf ~B27} (1997) 1719.
\bibitem{giudice} N. Arkani-Hamed et al., \PL {\bf B429} (1998) 263,
 hep-ph/9807344; 
G. F. Giudice et al., hep-ph/9811291; J. L. Hewett, hep-ph/9811356; 
S. Nussinov and R. Shrock, hep-ph/9811323.
\bibitem{dienes} K. R. Dienes et al., hep-ph/9807522; 
see also K. R. Dienes \PRep ~{\bf 287} (1997) 447.
\bibitem{grant} A. K. Grant, \PR {\bf ~D51} (1995) 207.
\bibitem{ckz} P. Chankowski, M. Krawczyk and J. \.Zochowski - in preparation.
\bibitem{santos} R. Santos and A. Barroso, \PR {\bf ~D56} (1997) 5366.
\bibitem{large} G. F. Giudice and Ridolfi, \ZP {\bf C41} (1988) 447; 
M. Olechowski and S. Pokorski, \PL {\bf B214} (1988) 393; A.Buras et al., 
\NP {\bf B271} (1985)44.
\bibitem{nir} V. Barger et al., \PR {\bf D41} (1990) 3421; 
Y. Grossman, \NP {\bf ~B426} (1994) 355.
\bibitem{sinbal3}The L3 Coll., M. Acciarri et al., \ZP {\bf C62} (1994) 551; 
submission to ICHEP'96 (Warsaw) PA11-016.
\bibitem{yukaleph}The ALEPH Coll., submitted to ICHEP'96, Warsaw, PA13-027.
\bibitem{TEV} B. Bevensee, FERMILAB-Conf-98-155-E, Moriond'98. 
\bibitem{sola}J. A. Coarosa et al., hep-ph/9808278.
\bibitem{mkz} a) M. Krawczyk, P. M\"{a}ttig and J. \.Zochowski,
 hep-ph/9811256; b)  ``2HDM at LC: $Z\ra h(A) + \gamma$'', 
 ECFA-DESY LC Meeting at Frascati, Nov. 1998.
\bibitem{kane} {\em Perspectives on Higgs physics}II, ed. G. Kane, 1997
(World Sci. Publ.).
\bibitem{carena}
M. Carena et.al., \PL {\bf ~B355} (1995) 209; M. Carena et al., 
\NP {\bf ~B461} (1996) 407; H. Haber et al., \ZP {\bf C75} (1997) 539;
S. Heinemeyer et al., \PR {\bf ~D58} (1998) 091701, (hep-ph/9807423).
\bibitem{marcela} M. Carena and C. Wagner, hep-ph/9808312;
C. Balazs et al. hep-ph/9807349 M. Spira, hep-ph/9810289.
\bibitem{opal366} OPAL Coll., CERN-EP-98/173.
\bibitem{drees} M. Drees et al., \PRL {\bf ~80} (1998) 2047, 
Erratum \PRL {\bf ~81} (1998) 2394.
\bibitem{roco} M. Roco and A. Belyaev, Higgs Working Group Meeting on Run II,
FERMILAB,   Nov. 1998. 
\bibitem{barger} V. Barger, hep-ph/9708442.
\bibitem{was} D. Froidevaux et al., ATLAS Internal Note, PHYS-N0-74 (1995).
\bibitem{nie} S. Nie and M. Sher, hep-ph/9811234.
\bibitem{g2} a) J. Bailey et al., \PL {\bf B68} (1977) 191; 
F.J.M. Farley and E. Picasso, \ARNS {\bf 29} (1979) 243;
F.J.M. Farley, \ZP {\bf C56} (1992) S88;\\
b) E 821 Coll., C. Timmermans, talk at ICHEP'98, Vancouver;
 A. Czarnecki, W. Marciano,hep-ph/9810512,  \\
c) M. Krawczyk and J. \.Zochowski, \PR {\bf ~D55} (1997) 6968.
\bibitem{mkk} M. Krawczyk, in proc. Workshop on Physics at
 the First Muon Collider, FERMILAB, Nov. 1997 (hep-ph/9803484).
\bibitem{greub}F.M. Borzumati and C. Greub, hep-ph/9802391;
C. Greub talk at ICHEP'98, Vancouver (hep-ph/9810240v2).\\
 ALEPH Coll., R. Barate et al., \PL {\bf B429} (1998) 169;
 CLEO Coll., submission to ICHEP'98, Vancouver.
\bibitem{brzhag}
 ALEPH Coll., R. Barate et al., \EPJ {\bf C4} (1998) 571; 
 DELPHI Coll., J.A. Barrio et al., internal note
         DELPHI 95-73 PHYS 508, submitted to the EPS-HEP conference '95;
 L3 Coll., M. Acciarri et al., \PL {\bf B388} (1996) 409;
 OPAL Coll., G. Alexander et al., \ZP {\bf C71} (1997) 1.
submission to ICHEP'96 (Warsaw) PA11-016.
\bibitem{gpr} B. Grzadkowski et al., \PL {\bf B272} (1991) 174.
\bibitem{bk}  A. Bawa and M. Krawczyk,IFT 16/91, 16/92;
\PL {\bf B357} (1995) 637;
\bibitem{mk}  M. Krawczyk, `` Higgs Search at HERA'', proc. ``
Future Physics at HERA'', 1995-96, DESY, eds. 
G. Ingelman, A. de Roeck, R. Klanner, p. 244.  
\bibitem{lowtest} D. L. Borden,  in proc. of LC Workshop 1993, p. 323; 
E. L. Saldin et al. \NIM {\bf ~A355} (1995) 171.  
\bibitem{lowlc} D. Choudhury and M. Krawczyk, \PR {\bf ~D55} (1997) 2774,
and in preparation.
\end{thebibliography}
\end{document}